\documentclass[aps,twocolumn,showpacs,amsmath,superscriptaddress,amssymb,prc]{revtex4-2}
\usepackage{graphicx,here}
\usepackage{multirow}
\usepackage{tikz}
\usepackage{epstopdf}
\usepackage[percent]{overpic}
\usepackage{scrextend}
\usepackage{xcolor,xspace}
\usepackage[ulem=normalem]{changes}
\usepackage{hyperref}
\hypersetup{colorlinks=True,urlcolor=blue,linkcolor=blue,citecolor=blue,filecolor=black}

\colorlet{Changes@Color}{red}
\usepackage{dcolumn,color,footnote,bm,braket}
\usepackage{url,longtable,tabularx}
\usepackage{threeparttable}

\usepackage{fancybox}
\usetikzlibrary{matrix}

\begin{document}

\title{Shape transition and coexistence in Te isotopes studied with the quadrupole collective Hamiltonian based on a relativistic energy density functional}

\author{K. Suzuki}
\affiliation{Department of Physics, 
Hokkaido University, Sapporo 060-0810, Japan}

\author{K. Nomura}
\email{nomura@sci.hokudai.ac.jp}
\affiliation{Department of Physics, 
Hokkaido University, Sapporo 060-0810, Japan}
\affiliation{Nuclear Reaction Data Center, 
Hokkaido University, Sapporo 060-0810, Japan}

\date{\today}

\begin{abstract}
Evolution and coexistence of shape and 
the related spectroscopic properties of even-even 
Te isotopes are investigated within the quadrupole 
collective model that is based on the 
nuclear density functional theory. 
By means of the constrained self-consistent mean-field 
calculations performed 
within the relativistic Hartree-Bogoliubov 
method with a choice of the energy density functional 
and pairing interaction, the 
deformation-dependent mass parameters and moments 
of inertia as well as collective potential of the 
triaxial quadrupole collective Hamiltonian 
are completely determined. 
The collective model produces for the near 
mid-shell nuclei, e.g., 
$^{116}$Te and $^{118}$Te, the low-energy 
$0^+_2$ state, which can 
be interpreted as the intruder state originating from 
the strongly deformed 
prolate minimum in the potential energy surface, 
along with the $0^+_1$ ground 
state that is attributed to the normal state 
based on a weakly oblate deformed global minimum. 
The collective model calculation suggests a parabolic 
behavior of the $0^+_2$ energy level 
near the neutron mid-shell $N=66$, 
as observed experimentally. 
Sensitivities of the calculated low-energy 
spectra to the pairing strength and collective 
mass parameters are analyzed. 
\end{abstract}

\maketitle

\section{Introduction}

The atomic nucleus is a composite system 
of neutrons and protons, where a subtle 
interplay between single-particle 
and collective motions plays a 
crucial role. 
Remarkable features of the nucleus are 
the fact that it organizes itself into a variety 
of shapes, and that the related collective 
excitations such as the anharmonic vibrations of the 
nuclear surface and rotational motions emerge \cite{BM}. 
The dominant nuclear 
deformation is of quadrupole type, 
which is characterized by the axial deformation 
$\beta$, i.e., degree of elongation 
along the symmetry axis of an ellipsoid, 
and the nonaxial or triaxial deformation $\gamma$, 
representing the deviation from the 
axial symmetry. 
Different values of the $\gamma$ deformation 
correspond to the prolate ($\gamma=0^{\circ}$), 
oblate ($\gamma=60^{\circ}$), and 
$\gamma$-soft ($0^{\circ}<\gamma<60^{\circ}$) shapes.

The nuclear structure undergoes a 
transition from a nearly spherical vibrational 
to deformed rotational states when it is seen as a function 
of the neutron number $N$ or proton 
$Z$ number. 
Such a nuclear structural evolution often 
takes place quite rapidly at a particular 
nucleon number, and is considered 
a quantum phase transition \cite{cejnar2010}, which 
is observable from striking 
differences of the measured intrinsic and 
spectroscopic properties between 
neighboring nuclei. 
In addition, several different nuclear shapes 
can coexist especially for those nuclei 
near the magic numbers, and several excited 
$0^+$ levels appear that are close in energy to 
the ground state \cite{heyde1992,wood1992,heyde2011,garrett2022}. 
The phenomenon of shape coexistence 
was found experimentally 
in the light mass region, i.e., $^{16}$O \cite{morinaga1956}, 
and later was identified in heavy nuclei 
as well, a representative example being 
the spherical-oblate-prolate triple 
shape coexistence in $^{186}$Pb \cite{andre00}, 
hence it is nowadays considered to be 
a universal feature observed 
in a wide spectrum of the chart 
of nuclides (see Ref.~\cite{garrett2022} 
for a recent review).

An interpretation of the shape coexistence  
is given by the spherical shell model: 
particularly for those nuclei that are near shell closure, 
multiparticle-multihole excitations are likely to occur, 
and they enhance correlations between valence neutrons 
and protons, which can be strong enough to substantially 
lower the non-yrast $0^+$ states to be in the 
vicinity of the ground state $0^+_1$ 
\cite{federman77,heyde85,heyde1992,wood1992,heyde95,heyde2011}. 
The different shell-model configurations are in the 
mean-field language related to 
several minima of the potential energy surface 
defined in the relevant collective coordinates. 
In the mean-field model calculations, 
several local minima that are 
rather close in energy to the global minimum 
often appear, and they are considered a signature of shape coexistence 
\cite{bengtsson1987,bengtsson89,wood1992,andre00,cwiok2005,heyde2011}.

Evidence for shape coexistence and the well established 
intruder bands built on the low-lying excited $0^+_2$ state 
are most spectacularly observed in those nuclei in the 
neutron-deficient Hg-Pb regions, where the proton 
two-quasiparticle $\pi (h_{9/2})^{2}$ 
and four-quasiparticle $\pi (h_{9/2})^{4}$ excitations 
are likely to occur near the mid-shell $N=104$. 
A similar mechanism could play a role in 
the low-lying states of 
those isotopes near the $Z=50$ major shall closure, 
including Cd ($Z=48$), Sn ($Z=50$), Te ($Z=52$), 
etc. isotopes. 
These nuclei were regarded as a textbook example 
for the anharmonic vibrator based on 
the multiphonon picture. 
In the case of the Cd isotopes, however, 
more recent experiments revealed 
additional low-spin levels near 
the multiphonon multiplets, 
and these new state could not be accounted for without 
introducing the contributions of the 
shape coexistence or intruder states 
coming from the next major oscillator shell. 
In the Sn as well as Cd isotopes, 
much experimental evidence has been found 
for the presence of 
the low-lying intruder band built on the $0^+_2$ state 
and its parabolic tendency toward the 
neutron mid-shell $N=66$ 
(see Ref.~\cite{garrett2022}, and references are therein).

It is then expected that the shape coexistence 
may play a role in Te isotopes with $Z=52$, 
as they are also close to the $Z=50$ 
magic number and the cross-shell excitations 
of the intruder orbitals could be relevant in low-lying states. 
Existence of intruder states 
in low-energy spectra of the even-even 
$^{116-128}$Te nuclei was pointed out by 
an earlier experimental 
study of Ref.~\cite{rikovska1989}. 
Lifetime measurements were performed 
for revealing a detailed level structure 
of $^{118}$Te \cite{mihai2011}. 
In recent years more and more 
experimental studies have been made 
to identify the 
intruder states and shape coexistence in 
near mid-shell Te isotopes. 
An experiment at the Cologne FN Tandem 
accelerator found several new 
low-energy levels in $^{116}$Te \cite{vonspee2024}. 
The latter two experimental studies, 
i.e., \cite{mihai2011,vonspee2024}, point to 
the possible occurrence of 
the shape coexistence in the mid-shell Te 
isotopes, whereas the theoretical interpretation 
used in those references was based on a 
simple interacting boson model (IBM) 
\cite{IBM} calculation, which did not 
take into account contributions from the intruder 
configuration, and in which the model 
parameters were fit to data. 
In addition, a lifetime measurement 
has recently been made investigating 
the $2^+_1$ energy and $B(E2;0^+_1 \to 2^+_1)$ 
transition of $^{118}$Te 
\cite{cederlof2023-118Te}. 
The same quantities for the $^{116}$Te and $^{118}$Te 
nuclei, together with their systematic trends 
along the entire Te isotopic chain in the 
neutron $N=50-82$ major shell were studied 
through another 
lifetime measurement, in comparison 
with the large-scale 
shell model calculations \cite{cbli2024}. 
Several other experiments on the Te nuclei 
with $N\geqslant 68$ have been performed 
in recent decades. 
For further details and 
the relevant references, the reader is 
referred to Ref.~\cite{garrett2022}. 
Much more experimental information is still 
required, in order to identify the evidence for 
shape coexistence and intruder bands in Te isotopes 
as firmly as in the case of the Sn and Cd isotopes, 
and this also necessitates systematic theoretical 
studies to be made in a timely manner.

From a theoretical point of view, 
low-energy excitation 
spectra and electromagnetic transition rates that 
signal the shape coexistence in the Te isotopes 
in open shell region have been carried 
out from various perspectives, 
e.g., the nuclear shell model 
\cite{qi2016,kaneko2021,sharma2022,cbli2024}, 
the static \cite{sharma2019,bonatsos2022} 
and beyond \cite{prochniak1999,libert2007,delaroche2010,CEA} 
self-consistent mean-field (SCMF) models 
based on the nuclear energy density functional (EDF), 
the IBM 
\cite{sambataro1982,rikovska1989,Lehmann97,pascu2010,mihai2011,gupta2023,vonspee2024}, the geometrical collective model \cite{budaca2021}
and an effective field theory \cite{coelloparez2015}. 
The SCMF method, in particular, 
that employs a given nonrelativistic 
(e.g., Skyrme \cite{Skyrme,bender2003,schunck2019} and 
Gogny \cite{Gogny,robledo2019}) or relativistic 
\cite{vretenar2005,niksic2011}
EDF is among the most reliable theoretical 
tools able to give a universal description of the 
nuclear matter properties, 
the intrinsic properties of finite nuclei including 
masses, radii, and deformations, 
and the low-energy collective excitations. 
To compute energies and wave functions of 
the excited states, the EDF framework 
needs to be extended to include 
dynamical correlations that are beyond 
the mean-field approximations, 
which are associated 
with the broken symmetries and with 
the quantum fluctuations around the 
mean-field minimum. 
It is then necessary to project 
mean-field solutions onto states with 
good symmetry quantum numbers and perform 
configuration mixing \cite{RS}. 
The beyond-mean-field 
treatment is made in principle by means of 
the generator coordinate method (GCM) 
\cite{RS,bender2003,robledo2019}. 
The GCM approach is, however, often 
computationally demanding for heavy nuclei, 
especially when the quadrupole triaxial or other 
higher-order shape degrees of freedom 
should be considered as collective coordinates. 
More computationally feasible EDF-based 
approaches to nuclear collective excitations 
are provided by solving the five-dimensional 
quadrupole collective Hamiltonian (QCH) 
\cite{niksic2009,delaroche2010,niksic2011}, 
and by mapping the mean-field potential 
energy surface (PES) onto the 
IBM Hamiltonian \cite{nomura2008,nomura2010,nomura2011rot}.

In the present article, by using the 
theoretical framework of the QCH that is 
based on the relativistic EDF, we study 
the low-energy collective states along the 
Te isotopic chain, and investigate possible 
shape coexistence around the neutron 
mid-shell $N=66$. 
Here the parameters of the QCH are 
completely determined 
by the constrained SCMF calculation employing 
given relativistic EDFs and pairing interactions. 
Over the last two decades the QCH within the relativistic EDF 
framework \cite{niksic2009,li2009,niksic2011} 
has found a number successful applications 
in describing various nuclear properties in 
different mass regions, including  
the shape coexistence 
(see, e.g., \cite{li2016,xiang2018,yang2021,nomura2022qch}), 
and quantum phase transitions 
(see, e.g., \cite{li2009,DDPC1-N28,niksic2011,DDPC1-SHE,konst2022}), 
and is thus considered a sensible microscopic 
approach to nuclear low-lying collective states. 
In particular, 
the same approach has recently been 
applied to study 
a shape phase transition and critical-point 
symmetry in the Kr isotopes below $N=82$ 
\cite{konst2022}, and the shape coexistence 
and intruder states in the Cd isotopes \cite{nomura2022qch}. 
Here we extend 
the previous EDF-based QCH calculation 
for the Cd isotopes \cite{nomura2022qch} 
to study low-lying states and physical quantities 
that indicate the relevance 
of shape coexistence along the chain of the 
Te isotopes, and give a timely theoretical 
description of the spectroscopic properties of 
those Te nuclei that are becoming 
of much interest experimentally. 
In addition, as a possible source of 
model uncertainty we further investigate the sensitivity of 
the results on relevant building blocks 
of the employed model, specifically, the 
pairing correlations considered in the 
SCMF model, and the collective mass parameters 
of the QCH.

This article is organized as follows. 
We briefly describe 
our theoretical framework in Sec.~\ref{sec:theory}. 
The mean-field results 
are presented in Sec.~\ref{sec:mf}, and 
the QCH results for the spectroscopic properties 
are discussed in Sec.~\ref{sec:results}. 
Section~\ref{sec:summary} gives a summary 
of the main results and concluding remarks.

\section{Theoretical framework\label{sec:theory}}

The first step of our theoretical analysis  
is to perform, for each Te nucleus, 
a set of the constrained SCMF calculations 
within the framework of the relativistic 
Hartree-Bogoliubov (RHB) method 
\cite{vretenar2005,niksic2011,DIRHB,DIRHBspeedup}. 
For the particle-hole channel, we employ two 
representative classes of the relativistic EDF: 
density-dependent point-coupling (DD-PC1) 
\cite{DDPC1} and density-dependent 
meson-exchange (DD-ME2)\cite{DDME2} interactions. 
Note, however, that since the results obtained from 
both EDFs are similar to each other, 
we mainly discuss in this paper  
the DD-PC1 results. 
For the particle-particle channel, 
we consider the separable 
pairing force of finite range that was 
developed in Ref.~\cite{tian2009}, given in 
coordinate space as
\begin{align}
\label{eq:pair1}
 V({\bf r}_{1},{\bf{r}_2},{\bf{r}'_1},{\bf{r}'_2})
=-V\delta({\bf R}-{\bf{R}'})P({\bf{r}})P({\bf{r}'})\frac{1}{2}(1-P^{\sigma}) \; ,
\end{align}
where ${\bf R}=({\bf r}_{1}+{\bf r}_{2})/2$ 
and ${\bf r}={\bf r}_{2} - {\bf r}_{2}$ are 
the center-of-mass and relative coordinates, respectively. 
The factor $P({\bf{r}})$ takes the form 
of a Gaussian function, 
\begin{align}
\label{eq:pair2}
 P({\bf{r}}) = \frac{1}{(4{\pi}a^2)^{3/2}}e^{-{\bf r}^{2}/4a^{2}} \; .
\end{align}
The strength parameter $V=728$ MeV fm$^{3}$ 
and the other parameter $a=0.644$ fm were fixed 
in Ref.~\cite{tian2009}, so that the $^{1}S_{0}$ pairing gap 
of infinite nuclear matter resulting from the 
Hartree-Fock-Bogoliubov (HFB) model calculation 
using the Gogny-D1S EDF \cite{D1S} should be reproduced.

The sensitivity of the mean-field results 
on the strength of the separable pairing force 
was systematically analyzed in Ref.~\cite{teeti2021}, 
and it was suggested that the pairing strength $V$ 
in Eq.~(\ref{eq:pair1})
may vary as functions of nucleon number 
in order to account for the empirical odd-even mass staggering 
in medium-heavy and heavy nuclei. 
$V$ should be subject to be modified 
as $V \to V'=fV$, with $f$ being 
a scaling factor. 
In the present study, we consider the default 
pairing strength with $f=1.0$ and the 
value increased by 5\% of $V$, corresponding to $f=1.05$. 
We further adopt a type of 
the factor that takes into account 
the neutron-proton number difference, i.e., 
\begin{eqnarray}
\label{eq:pair3}
 f(N) = c_1 \exp\left(c_2 \frac{|N-Z|}{A}\right) \; ,
\end{eqnarray}
with $c_1$ and $c_2$ being parameters. 
The above form (\ref{eq:pair3}) was considered in 
Ref.~\cite{teeti2021}, and was shown to gradually 
decrease along an isotopic chain. 
In all these three cases of the pairing strength, 
we consider equal values for the neutron and proton pairings, 
as in Ref.~\cite{tian2009}. 
The parameters $c_1=1.1$ and $c_2=-0.58$ are here 
chosen, so as to fulfill 
the conditions that they are more or less similar to 
those values typically considered in Ref.~\cite{teeti2021}, and 
that the factor $f$ varies along the 
isotopic chain from 1.1 ($N=52$) to 0.97 ($N=80$), 
leading to an overall agreement with the 
experimental $2^+_1$ level on both ends of the 
isotopic chain, i.e., at $N \approx 54$ and $N \approx 78$.

The constraints imposed in the SCMF calculations are 
on the expectation values of the 
mass quadrupole operators
\begin{align}
 \hat Q_{20}=2z^2-x^2-y^2 
\quad\text{and}\quad
 \hat Q_{22}=x^2-y^2 \; ,
\end{align}
which are related to the 
axially symmetric deformation $\beta$ 
and triaxiality $\gamma$ \cite{BM}, through 
the relations 
\begin{align}
\label{eq:bg}
& \beta=\sqrt{\frac{5}{16\pi}}\frac{4\pi}{3}\frac{1}{A(r_{0}A^{1/3})^{2}}
\sqrt{\braket{\hat{Q}_{20}}^{2}+2\braket{\hat{Q}_{22}}^{2}} \; , \\
& \gamma=\arctan{\sqrt{2}\frac{\braket{\hat{Q}_{22}}}{\braket{\hat{Q}_{20}}}} \; ,
\end{align}
with $r_0=1.2$ fm. 
The SCMF calculations are carried out 
in a harmonic oscillator basis, with the number of 
oscillator shells equal to 14.

Quadrupole collective states are 
obtained by solving the QCH, or the collective 
Schr\"odinger equation. 
The parameters 
of the QCH are completely determined by using 
the results of the RHB calculations: 
the quadrupole triaxial ($\beta,\gamma$) PES, 
and the single-particle solutions. 
More detailed accounts of the QCH 
are found, e.g., in Refs.~\cite{niksic2009,niksic2011}. 
The QCH, denoted as 
$\hat{H}_{\textnormal{coll}}$, 
is given as
\begin{align}
\label{eq:hamiltonian-quant}
\hat{H}_{\textnormal{coll}} 
= \hat{T}_{\textnormal{vib}}+\hat{T}_{\textnormal{rot}}
+V_{\textnormal{coll}} \; ,
\end{align}
with the vibrational kinetic energy
\begin{align}
\label{eq:vib}
\hat{T}_{\textnormal{vib}} =
&-\frac{\hbar^2}{2\sqrt{wr}}
   \Biggl[\frac{1}{\beta^4}
   \Biggl(\frac{\partial}{\partial\beta}\sqrt{\frac{r}{w}}\beta^4
   B_{\gamma\gamma} \frac{\partial}{\partial\beta}
   \nonumber \\
&-\frac{\partial}{\partial\beta}\sqrt{\frac{r}{w}}\beta^3
   B_{\beta\gamma}\frac{\partial}{\partial\gamma}
   \Biggr)
   +\frac{1}{\beta\sin{3\gamma}}\Biggl(
   -\frac{\partial}{\partial\gamma} \sqrt{\frac{r}{w}}\sin{3\gamma}
\nonumber \\
&\times
B_{\beta \gamma}\frac{\partial}{\partial\beta}
    +\frac{1}{\beta}\frac{\partial}{\partial\gamma} \sqrt{\frac{r}{w}}\sin{3\gamma}
      B_{\beta \beta}\frac{\partial}{\partial\gamma}
   \Biggr)\Biggr] \; ,
\end{align}
rotational kinetic energy
\begin{align}
\label{eq:rot}
\hat{T}_{\textnormal{rot}} =
\frac{1}{2}\sum_{k=1}^3{\frac{\hat{J}^2_k}{\mathcal{I}_k}} \; ,
\end{align}
and collective potential 
$V_{\textnormal{coll}}(\beta,\gamma)$. 
Note that the operator $\hat{J}_k$ in Eq.~(\ref{eq:rot}) 
denotes the components of the angular momentum in
the body-fixed frame of a nucleus. 
The mass parameters 
$B_{\beta\beta}$, $B_{\beta\gamma}$, 
and $B_{\gamma\gamma}$ in Eq.~(\ref{eq:vib})
and the moments of inertia (denoted as MOIs)
$\mathcal{I}_k$ in (\ref{eq:rot}) 
are functions of the $\beta$ and $\gamma$ 
deformations, 
and are related to each other by 
the quantity 
$\mathcal{I}_k = 4B_k\beta^2\sin^2(\gamma-2k\pi/3)$. 
Two additional quantities in Eq.~(\ref{eq:vib}), i.e., 
$r=B_1B_2B_3$, and $w=B_{\beta\beta}B_{\gamma\gamma}-B_{\beta\gamma}^2 $,
determine the volume element in the collective space. 
The moments of inertia are
computed using the Inglis-Belyaev (IB) formula 
\cite{inglis1956,belyaev1961}, and the mass
parameters are calculated in the cranking 
approximation. 
A well-known fact is that the IB formula 
considerably underestimates the empirical 
moments of inertia, and in order 
to account for the discrepancy it is often 
increased as 
\begin{eqnarray}
\label{eq:moi}
 \mathcal{I}_k' = (1+\alpha)\mathcal{I}_k \; ,
\end{eqnarray}
with the scaling factor being typically 0.3--0.4 
in literature. 
In this study, we scale the IB moment of inertia 
with a factor $\alpha=0.3$ for all the Te nuclei. 

The collective potential 
$V_{\textnormal{coll}}(\beta,\gamma)$ 
(\ref{eq:hamiltonian-quant}) is obtained by 
subtracting the zero-point energy corrections 
from the total RHB deformation energy.

The corresponding eigenvalue problem is solved using 
an expansion of eigenfunctions in terms
of a complete set of basis functions that depend on the 
deformation variables $\beta$ and
$\gamma$, and the three Euler angles. 
The diagonalization of the QCH with the ingredients 
specified by the RHB-SCMF calculation 
yields the excitation energies and collective
wave functions for each value of the total angular 
momentum and parity, 
that are used to calculate various physical observables. 
A virtue of using the QCH 
based on SCMF single-(quasi)particle 
solutions is the fact that the electromagnetic 
transition properties, such as the 
electric quadrupole ($E2$) 
and monopole ($E0$) transition rates, 
and spectroscopic quadrupole 
moments, are calculated in the full configuration 
space, hence there is no need for effective charges. 
Using the bare value of the proton charge in the
electric transition operators, 
the transition probabilities 
between eigenvectors of the QCH  
can be directly compared with 
spectroscopic data.

\begin{figure*}
\begin{center}
\includegraphics[width=0.7\linewidth]{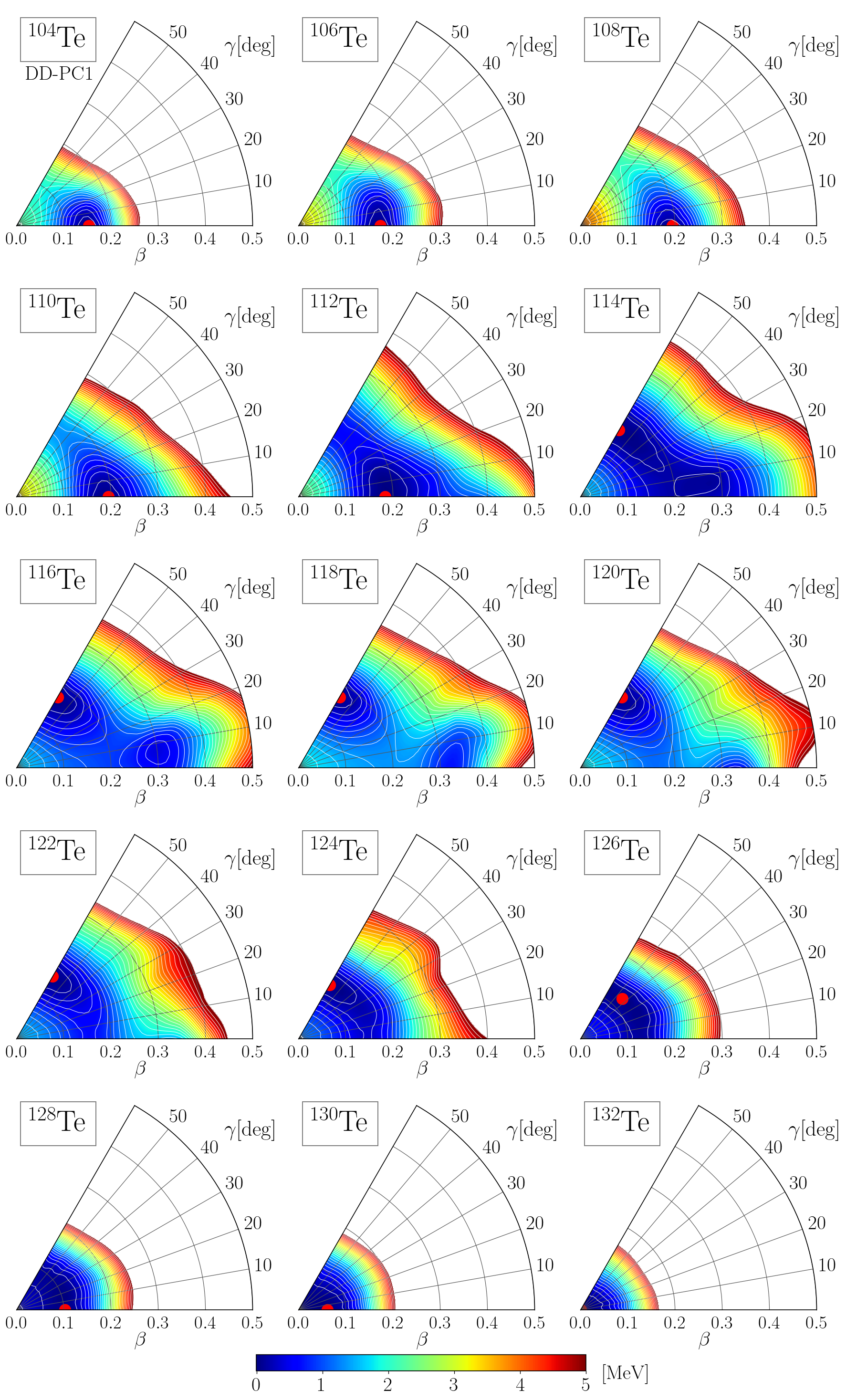}
\caption{Potential energy surfaces as functions of the 
triaxial quadrupole ($\beta,\gamma$) deformations 
for the even-even nuclei $^{104-132}$Te 
computed with the constrained 
relativistic Hartree-Bogoliubov 
method employing the DD-PC1 energy density 
functional and the separable pairing force of 
finite range. The default pairing strength 
$V$ ($f=1.0$) is used. 
}
\label{fig:pes}
\end{center}
\end{figure*}

%
%
\begin{figure*}
\begin{center}
\includegraphics[width=.7\linewidth]{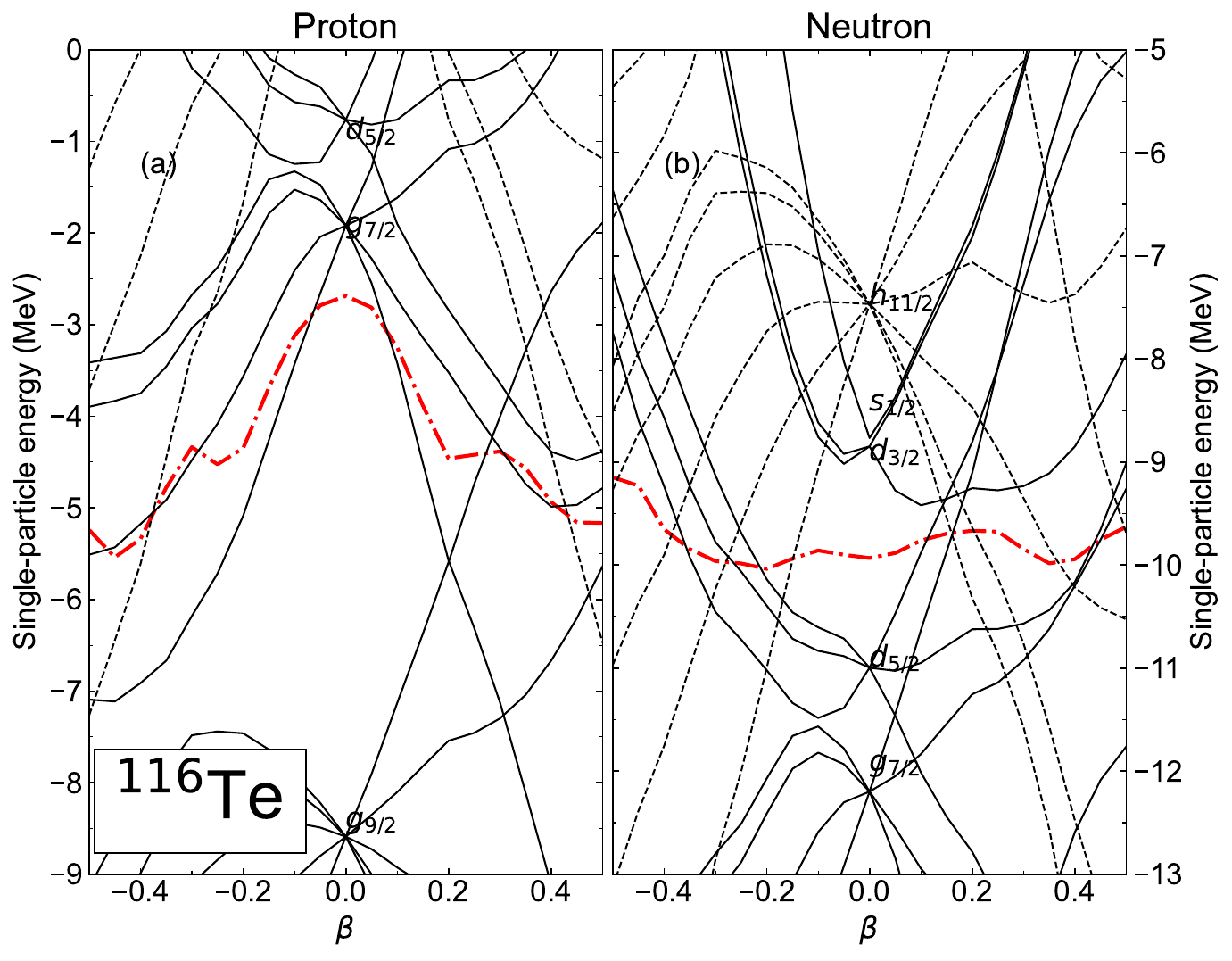}
\caption{
Single-particle energies (in MeV) for protons (left) 
and neutrons (right) for $^{116}$Te as functions 
of the axial quadruple deformation $\beta$. 
The dashed-dotted curves stand for the Fermi energies. 
}
\label{fig:nilsson}
\end{center}
\end{figure*}

%
%
\begin{figure*}
\begin{center}
\includegraphics[width=.6\linewidth]{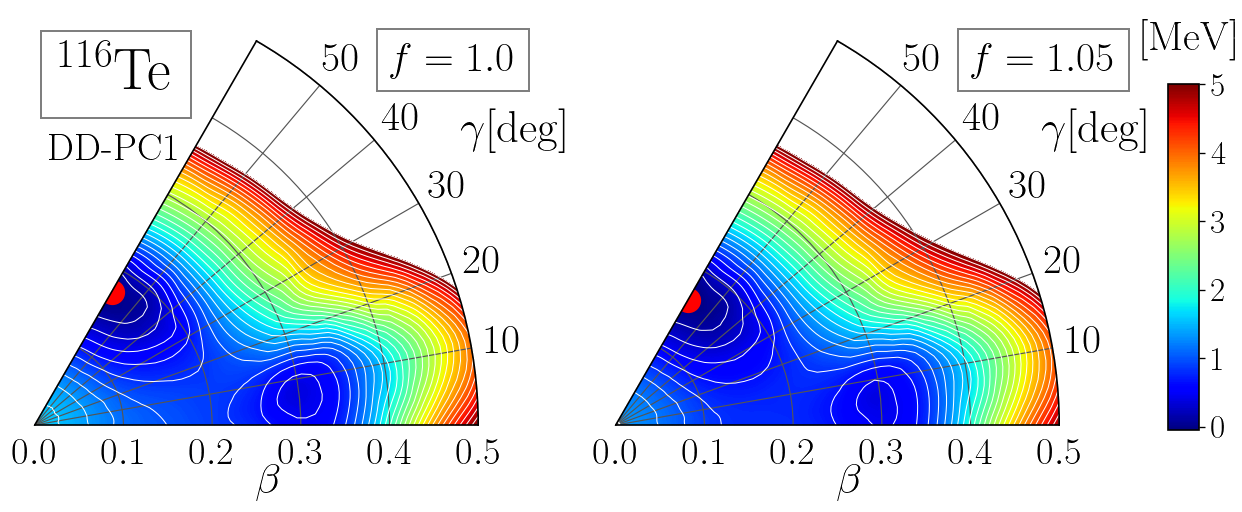}
\caption{
$(\beta,\gamma)$ PESs for $^{116}$Te 
obtained from the RHB-SCMF calculations 
using the default pairing strength $V=728$ MeV fm$^3$ 
with a factor $f=1.0$ (left), 
and the increased strength 
$1.05V=764$ MeV fm$^3$ with $f=1.05$ (right). 
The DD-PC1 EDF is used. 
}
\label{fig:pes-pairing}
\end{center}
\end{figure*}

\begin{figure}[ht]
\begin{center}
\includegraphics[width=\linewidth]{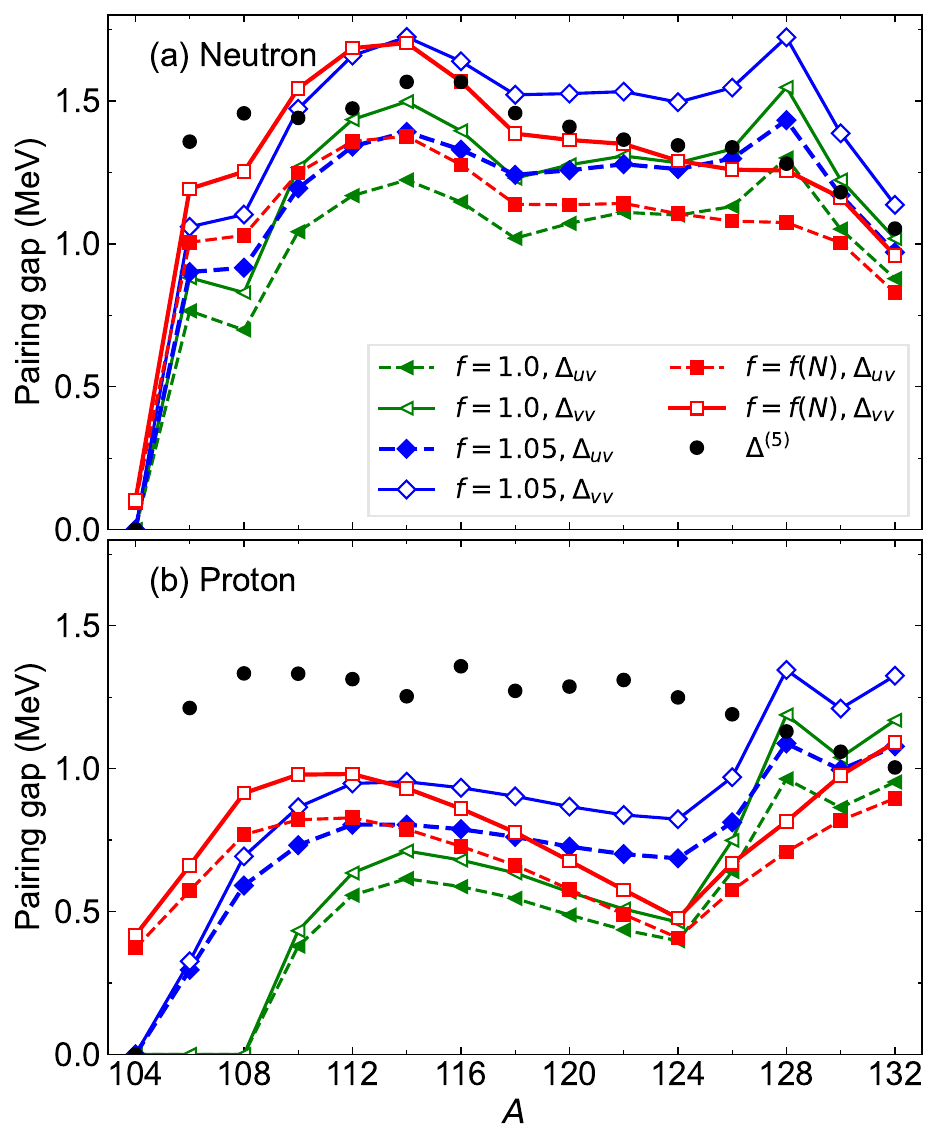}
\caption{
Calculated (a) neutron and (b) proton pairing gaps 
$\Delta_{uv}$ (\ref{eq:uv}) and $\Delta_{vv}$ (\ref{eq:vv})
in the cases of the three different 
pairing strengths corresponding to the scaling factors 
$f=1.0$, $f=1.05$, and $f=f(N)$ [Eq.~(\ref{eq:pair3})] 
for the studied $^{104-132}$Te nuclei. 
$\Delta^{(5)}$ represents the empirical pairing gap 
in the five-point formula 
calculated by using the experimental binding energies 
available at AME2020 \cite{huang2021,wang2021}. 
}
\label{fig:gap}
\end{center}
\end{figure}

\section{Mean-field results\label{sec:mf}}

\subsection{Potential energy surfaces}

Figure~\ref{fig:pes} shows the quadrupole triaxial 
PESs for the studied even-even $^{104-132}$Te 
isotopes calculated by the constrained SCMF 
method within the RHB model that is 
based on the DD-PC1 EDF and separable pairing 
interaction (\ref{eq:pair1}) 
described in the previous section. 
As can be seen in Fig.~\ref{fig:pes}, 
for light Te isotopes, $^{104-110}$Te, with 
the neutron numbers $N=52-58$, the potential has 
the absolute minimum on the prolate axis 
around $\beta=0.15$. 
For those Te nuclei heavier than $^{110}$Te, 
the PES becomes considerably 
softer in $\gamma$ deformation, 
indicating the strong shape mixing 
with respect to the nonaxial quadrupole deformation. 
A notably $\gamma$-soft energy surface 
with a shallow oblate global minimum at $\beta \approx 0.15$ 
is obtained for the $^{114}$Te nucleus. 
In addition to the global minimum on the 
oblate side, a nearly prolate, 
triaxial local minimum 
becomes visible in the $^{116,118,120}$Te nuclei. 
This points to the onset of the oblate-prolate 
shape coexistence 
toward the middle of the neutron major 
shell, i.e., $N=66$. 
From $^{122}$Te and the heavier Te nuclei, however, 
the local minimum on the prolate 
side diminishes, and 
only a weakly oblate deformed 
ground state is suggested. 
Approaching $^{130}$Te, 
the minimum shifts to the spherical configuration, 
$\beta=0$, reflecting the neutron shell closure $N=82$. 

The non-relativistic Gogny 
Hartree-Fock-Bogoliubov (HFB) calculation 
using the D1S force \cite{CEA,delaroche2010} 
predicts a similar systematic behavior 
of the intrinsic shape along the Te isotopic chain. 
A major difference between the present and 
the Gogny-HFB calculations is that the 
latter gives the global mean-field minimum 
on the prolate side for $^{116,118,120}$Te, 
and both the prolate global minimum and 
oblate local minimum for these nuclei are 
quite close in energy. 
The present calculation, however, 
predicts an oblate global minimum, along with 
the prolate local minimum, as mentioned above. 
The SCMF calculations have been reported 
in Ref.~\cite{sharma2019}, which also adopted 
the RHB model but with the DD-ME2 EDF, 
in addition to the DD-PC1 EDF. 
In that work, mainly the 
systematic of the deformations and other 
intrinsic properties of the Te isotopic chain
were studied, and the ($\beta,\gamma$) PESs obtained 
there for the Te nuclei are 
quite similar to those in the 
present calculation. 

\subsection{Single-particle energy spectra}

In addition to the topology of the mean-field PES, 
the onset of deformation, especially that of 
shape coexistence, can be inferred from the behaviors of 
the single-particle spectra as functions of 
deformation, or Nilsson-like plots, 
which are obtained from the RHB calculation with 
the constraint on axial $\beta$ deformation. 
Should a gap appear  
in the regions in which the density 
of the single-particle levels near the Fermi 
energies is low, it implies 
stability in nuclear structure 
at those nucleon numbers that are different 
from the spherical magic numbers. 
This situation is actually observed in 
Fig.~\ref{fig:nilsson}, in which 
the proton and neutron single-particle 
energies near the Fermi energies are shown 
as functions of the axial quadrupole 
deformation $\beta$ for $^{116}$Te. 
At the mean-field level the oblate-prolate shape coexistence 
is particularly pronounced in this nucleus 
(cf. Fig.~\ref{fig:pes}). 
In the proton single-particle spectra shown in 
Fig.~\ref{fig:nilsson}(a), a large gap between 
the $\pi g_{7/2}$ and $\pi g_{9/2}$ 
orbitals at $\beta=0$ represents the $Z=50$ 
major shell closure. 
There is a region on the oblate ($\beta \approx -0.2$) 
side, in which level density near the 
proton Fermi energy is low. 
This can be related to the oblate ground state 
in the corresponding ($\beta,\gamma$) PES. 
In the neutron single-particle spectra, 
depicted in Fig.~\ref{fig:nilsson}(b), 
one can notice a gap 
between the levels coming from the $\nu d_{3/2}$ 
and $\nu d_{5/2}$ single-particle orbitals, 
which is mainly extended to the oblate region 
$-0.2 \lesssim \beta \lesssim 0.1$. 
The neutron level density is also low at $\beta \approx 0.3$, 
which is created by the states 
originating from the $\nu h_{11/2}$, 
$\nu d_{3/2}$, $\nu d_{5/2}$, and 
$\nu g_{7/2}$ spherical single-particle states. 
The appearance of the gap in the neutron 
single-particle plot around the region 
$\beta \approx 0.3$ near the Fermi energy 
then implies that the prolate local 
minimum may occur as a consequence of the 
lowering of the intruding orbitals from the 
spherical $\nu h_{11/2}$ orbital. 
A similar, but much less pronounced 
gap on the prolate side is 
also visible in the proton single-particle 
spectra, which is formed by the lowering 
of the levels coming from the $\pi h_{11/2}$ 
orbital.

\subsection{Dependencies on the pairing strength}

We compare in Fig.~\ref{fig:pes-pairing} 
the PESs computed for $^{116}$Te 
by using the default value of the 
pairing strength 
$V=728$ MeV fm$^3$ (left) 
and the increased strength 
$1.05V=764$ MeV fm$^3$ (right), 
in the RHB-SCMF calculations 
based on the DD-PC1 EDF. 
For both nuclei, an impact of increasing 
the pairing strength in the RHB-SCMF calculation 
is such that the PES becomes softer in both 
$\beta$ and $\gamma$ deformations, and 
the prolate local minimum appears to be 
less pronounced in the calculation with the 
increased pairing strength. 
This reflects the fact that the pairing 
correlations rather prefer a less 
deformed nuclear shape. 
The increase of the pairing strength 
seems to have similar effects on the PESs 
for other Te nuclei, especially those 
near the mid-shell. The same is true 
when the DD-ME2 EDF is employed.

In addition to the potential energies, 
we study the sensitivity of the calculated pairing 
gaps to the pairing strength. 
Figure~\ref{fig:gap} shows the neutron and 
proton pairing gaps resulting from the RHB-SCMF calculations, 
performed at the $(\beta,\gamma)$ coordinates 
corresponding to the global minimum, 
in the three different choices of 
the pairing strength with the 
scaling factors $f=1.0$ (default value), $f=1.05$, and 
$f=f(N)$ [see Eq.~(\ref{eq:pair3})]. 
$\Delta^{(5)}$ plotted in the figure represents 
the empirical pairing gap in the 
five-point formula, which is calculated by using the 
experimental binding energies available in 
Atomic Mass Evaluation 2020 (AME2020) \cite{huang2021,wang2021}. 
To compute pairing gaps we employ the following formulas 
that are often used in the literature: 
\begin{eqnarray}
\label{eq:uv}
 \Delta_{uv}=\frac{\sum_{k}u_kv_k \Delta_k}{\sum_{k}u_kv_k}
\end{eqnarray}
and
\begin{eqnarray}
\label{eq:vv}
 \Delta_{vv}=\frac{\sum_{k}v_k^2 \Delta_k}{\sum_{k}v_k^2} \; ,
\end{eqnarray}
where $u_k$ and $v_k$ are unoccupation and occupation 
amplitudes, respectively, $\Delta_k$ is the diagonal 
matrix element of the pairing field, and the sums in both 
equations run over all the considered states $k$ 
in the canonical basis. See, e.g., Ref.~\cite{agbemava2014}, 
for the details about the above formulas. 

We observe from Fig.~\ref{fig:gap}(a) that 
the calculated neutron gap is rather sensitive to 
the pairing strength used in the RHB-SCMF model. 
This feature appears to hold regardless of which of 
the formulas $\Delta_{uv}$ and $\Delta_{vv}$ is employed, 
and is well illustrated by the fact that 
in Fig.~\ref{fig:gap}(a) the 
calculated $\Delta$'s with the scaling factor $f=1.05$ 
are systematically larger than those with the 
default pairing strength with $f=1.0$ by typically 10--20 \%. 
By using the $N$-dependent scaling factor 
for the neutron pairing strength, the corresponding 
pairing gap in each of the
two formulas (\ref{eq:uv}) and (\ref{eq:vv}) 
turns out to be larger than that obtained 
with the default strength $f=1.0$ on the 
neutron-deficient side ($A\leqslant 112$), but is reduced  
on the neutron-rich side ($A\geqslant 124$). 
Among all the calculated neutron pairing gaps 
shown in Fig.~\ref{fig:gap}(a), the $\Delta_{vv}$ values 
calculated with the $N$-dependent scaling factor $f(N)$ 
are generally closest to the empirical neutron five-point 
pairing gaps, except perhaps for the $^{110,112,114}$Te nuclei.

The calculated proton pairing gaps shown in 
Fig.~\ref{fig:gap}(b) exhibit similar dependencies on 
the pairing strength to the neutron gaps, i.e., 
they increase (decrease) as the pairing strength 
is increased (decreased). 
As compared to the neutron gaps, the proton ones 
are predicted to be small since in most cases $\Delta<1$, 
but are also sensitive to the pairing strength especially 
for the neutron-deficient nuclei. 
One can observe that the 
calculated proton gaps significantly underestimate 
the empirical ones $\Delta^{(5)}$ for $A\lesssim 124$, 
and the disagreement becomes even more significant 
especially near the $N=50$ neutron shell closure. 
This discrepancy is perhaps related to the fact that since the 
proton number $Z=52$ of the Te isotopes is 
close to the magic number, where the occupation 
probabilities $v^2$ are close to either 0 or 1 near 
the Fermi surface, the formulas 
$\Delta_{uv}$ and $\Delta_{vv}$ do not necessarily 
give reasonable values. 
Another source of the discrepancy is that the same pairing 
strength is considered for both protons and neutrons 
in the present calculation, whereas the strength could be 
in principle different between the proton and neutron systems. 
To reproduce the empirical proton pairing gaps, 
the scaling factor for the proton pairing strength 
could be, for instance, much larger than the 
default value, $f\gg1$, as one approaches 
the neutron-deficient side. 
The use of different pairing strengths for protons 
and neutrons would affect the spectroscopic results, 
but in order not to complicate the discussions 
we keep the equal neutron and proton pairing strengths 
in the present study.

\begin{figure*}[ht]
\begin{center}
\includegraphics[width=\linewidth]{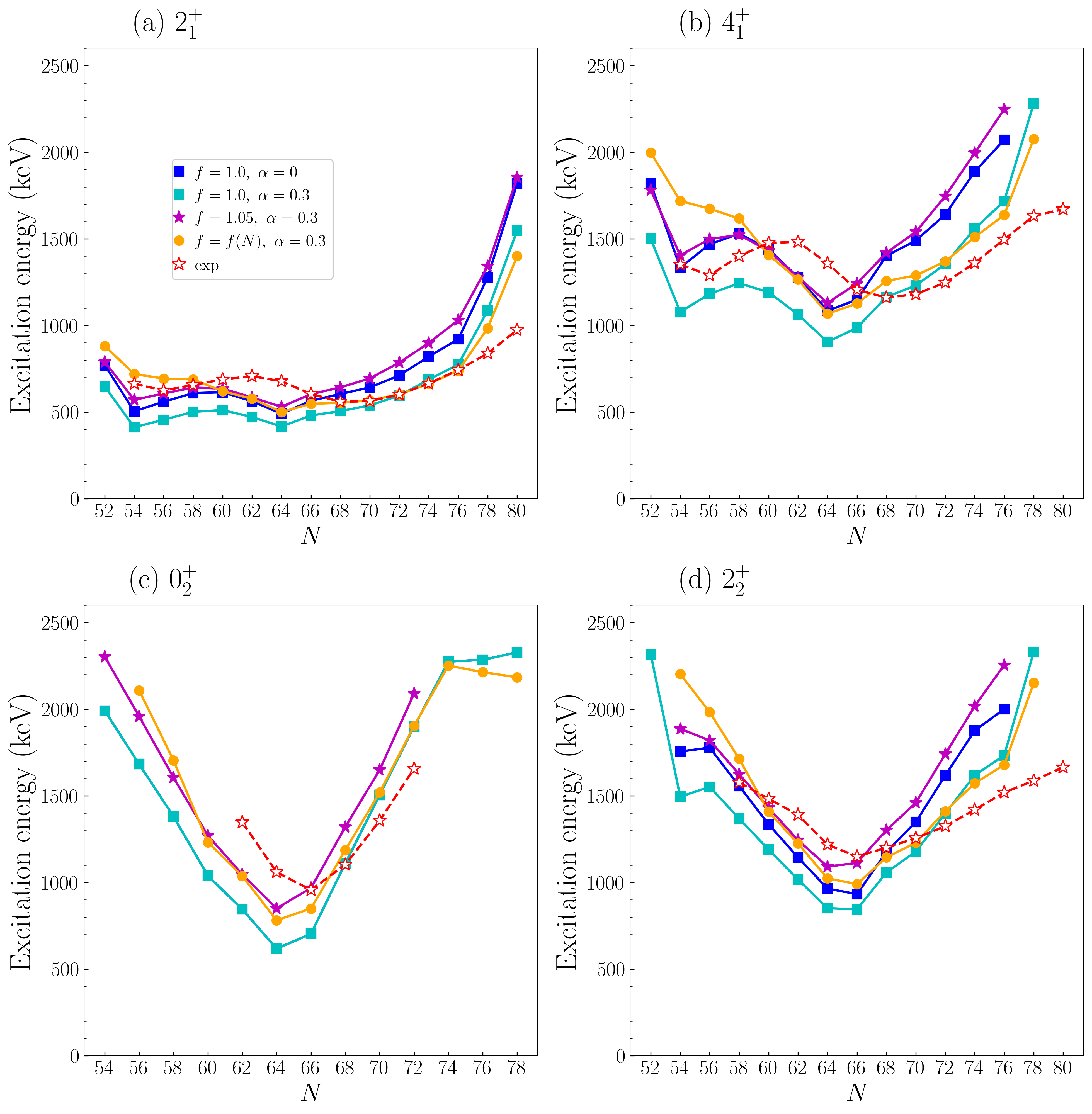}
\caption{
Calculated excitation energies for the (a) $2^+_1$, 
(b) $4^+_1$ , (c) $0^+_2$, and (d) $2^+_2$ states 
predicted by the RHB+QCH model. 
The experimental data are taken from Refs.~\cite{data,vonspee2024}. 
The DD-PC1 EDF is employed. 
In each panel, the calculated 
energies using the (i) default pairing and MOI 
($f=1.0, \alpha=0$), (ii) default pairing 
and increased MOI ($f=1.0, \alpha=0.3$), 
(iii) increased pairing and MOI ($f=1.05, \alpha=0.3$), 
and (iv) $N$-dependent [Eq.~(\ref{eq:pair3})] 
pairing strength and increased MOI [$f=f(N), \alpha=0.3$] 
are shown.
}
\label{fig:level}
\end{center}
\end{figure*}

\section{Spectroscopic results\label{sec:results}}

\subsection{Energy spectra}

We show in Fig.~\ref{fig:level} 
the calculated energy levels for the $2^+_1$, $4^+_1$, 
$0^+_2$, and $2^+_2$ states 
using the RHB+QCH method in the case of 
the DD-PC1 EDF, and compare with 
the experimental data available 
from Ref.~\cite{vonspee2024} 
and from the NNDC database \cite{data}. 
In each panel of Fig.~\ref{fig:level}, 
different RHB+QCH results are 
depicted, obtained from the calculations 
that employ 
(i) the original pairing strength $f=1.0$ 
and increased MOI with $\alpha=0.3$, 
(ii) the pairing strength increased by 5 \% ($f=1.05$) 
and increased MOI with $\alpha=0.3$, 
(iii) the nucleon-number dependent 
factor $f$ (\ref{eq:pair3}) and increased 
MOI with $\alpha=0.3$, and, as a reference,
(iv) the original pairing ($f=1.0$) 
as well as MOI ($\alpha=0$).  
As a general remark, 
increasing the MOI has an effect 
of lowering the excitation energies of the 
states with spins other than $0^+$, while 
the energy levels of the excited $0^+$ states 
remain the same. 
The increase in the pairing strength makes 
the whole energy spectrum stretched.

In Figs.~\ref{fig:level}(a) and \ref{fig:level}(b), 
it is seen that a characteristic behavior of the observed 
$2^+_1$ and $4^+_1$ energy levels as functions of the 
neutron number $N$ is an approximate parabola that 
is found at $N=56$ and 70. 
The RHB+QCH calculations 
with the fixed paring strengths (i.e., $f=1.0$ and $1.05$) 
for the entire isotopes provide similar systematic 
for the $2^+_1$ and $4^+_1$ levels, as they 
are lowest in energy at $N=64$, and are locally 
minimal at $N=54$. 
These neutron numbers are at variance 
with experiment. 
We note, however, that the EDF is usually 
not specifically adjusted to reproduce low-lying 
states along a given isotopic chain, 
and thus it is rather natural that 
the difference in a few nucleon numbers appears. 
The RHB+QCH predictions with the default 
pairing strength ($f=1.0$) 
underestimate the $2^+_1$ and $4^+_1$ 
levels particularly for $N \leqslant 68$. 
For those nuclei with $70 \leqslant N \leqslant 76$, 
a reasonable description of the energy 
spectra is obtained with the 
increased ($\alpha=0.3$) MOI. 
All the model calculations overestimate 
the observed $2^+_1$ energies from $N>76$ toward the 
neutron major shell closure $N=82$. 
Note, however, that the collective 
model description becomes less reliable 
for those nuclei near the magic numbers, where 
single-particle degrees of freedom come to 
play a significant role.

The RHB+QCH calculations with the increased 
pairing with the fixed factor $f=1.05$ 
and the MOI with $\alpha=0.3$, 
produce the $2^+_1$ and $4^+_1$ yrast levels 
that are lying rather high in energy as compared 
to those calculations with the default 
pairing $f=1.0$, and even higher excitation 
energies are obtained for $N\geqslant 68$ and 
as one approaches the $N=82$ shell closure. 
To improve the description of the systematic 
for $N>66$, it may be expected that 
the pairing strength should decrease as a function of $N$ 
so that the energy levels 
are effectively lowered to be closer to the 
experimental ones. 
It is seen in Figs.~\ref{fig:level}(a) and \ref{fig:level}(b) 
that the RHB+QCH calculation that employs the 
$N$-dependent pairing strength (\ref{eq:pair3}) 
does indeed give a slightly 
better description of the $2^+_1$ systematic 
for $N \geqslant 66$ and $N \leqslant 58$. 
The same calculation, however, significantly 
overestimates the $4^+_1$ energies for $N\leqslant 58$. 
This indicates that the employed pairing strengths 
for these nuclei may have been too large. 
It is worth mentioning that 
all the four RHB+QCH results give 
the $2^+_1$ and $4^+_1$ yrast levels 
for those nuclei with $N=60-66$ that are 
lower than the experimental values. 
This is probably the consequence of the 
fact that the present RHB-SCMF calculation for 
the near mid-shell nuclei overestimates 
the quadrupole deformation, that is, the 
potential is so steep that the 
resulting collective energy spectrum becomes 
rather rotational like. 
For the mid-shell nuclei the RHB-SCMF 
also predicts coexisting minima and 
considerable shape mixing, rendering 
the low-spin ($2^+_1$) state 
unexpectedly lower in energy due to 
the level repulsion effect. 
Note that the Gogny-D1S HFB plus five-dimensional 
collective Hamiltonian (5DCH) calculation using the 
Gaussian overlap approximation \cite{CEA} 
obtained the $2^+_1$ excitation energies 
of 336 and 368 keV for $^{116}$Te and $^{118}$Te, 
respectively, which are close to the 
present results with the default pairing 
strength $f=1.0$ and increased MOI ($\alpha=0.3$), 
shown in Fig.~\ref{fig:level}(a).

In Fig.~\ref{fig:level}(c), 
experimentally the $0^+_2$ energy level exhibits 
a parabolic behavior with the minimal energy 
at the middle of the major shell $N=66$. 
The nature of the $0^+_2$ states in the near mid-shell 
Te nuclei has not yet been firmly identified. 
One explanation is that it belongs to the 
two-phonon multiplets of an anharmonic vibrator, 
since this state lies close in 
energy to the $4^+_1$ and $2^+_2$ levels. 
Alternative interpretation may be that it originates 
from the particle-hole excitation from below the 
$Z=50$ proton major shell, which gives rise to 
a shape coexistence. 
The present RHB-QCH calculation seems to support 
the interpretation in terms of 
shape coexistence, based on the fact 
that an oblate global minimum and 
a well developed near prolate minimum appear 
in the triaxial quadrupole PESs (cf. Fig.~\ref{fig:pes}). 
As shown in the next section, 
the calculated $0^+_2$ states specifically for 
the $^{116,118}$Te nuclei 
in the neutron mid-shell $N=66$ are composed mainly 
of the mean-field configurations near the 
prolate secondary minimum. 
Overall, all the four RHB+QCH calculations 
give the $0^+_2$ excitation energies that are 
systematically much lower than the experimental ones 
for the mid-shell Te nuclei. 
%
The RHB-SCMF PESs corresponding to these nuclei 
suggest that both the prolate 
local and oblate global minima are rather close in 
energy to each, and this may explain 
the unexpectedly low-energy $0^+_2$ levels as compared 
to the experimental ones. 
%
In addition, the different calculations 
all suggest the lowest 
$0^+_2$ energy to occur at $N=64$, whereas experimentally 
the lowest $0^+_2$ level is observed rather at $N=66$. 
With the increased pairing strength with $f=1.05$ 
and the $N$-dependent strength, 
the $0^+_2$ level are pushed up. 
This change is most clearly seen for those 
nuclei with $N \leqslant 66$, 
where, for instance, a better agreement 
with data is observed for $^{118}$Te when the 
pairing strength with $f=1.05$ is employed.  
The calculation with the $N$-dependent 
pairing strength also gives a similar result 
for the $0^+_2$ energy level systematic 
to that with the increased strength $f=1.05$. 
The Gogny-D1S HFB plus 5DCH calculation of 
Ref.~\cite{CEA} obtained for $^{116}$Te and $^{118}$Te 
the $0^+_2$ excitation energies of 
755 and 750 keV, respectively, which are also 
similar to the present values.

Interpretation of the $2^+_2$ state is not trivial 
in the present calculation, since as will be 
shown both $K^\pi=0^+$ and $2^+$ components are 
significantly mixed in this state. 
It may be considered a member 
of the two-phonon multiplets, the bandhead of 
the $\gamma$-vibrational band, or the member of the 
excited $0^+$ band. 
The experimental systematic of this state, 
shown in Fig.~\ref{fig:level}(d), exhibits a 
parabolic pattern around $N=66$, similarly to the 
$0^+_2$ systematic. 
The present RHB+QCH calculations seem to 
produce a parabolic trend that is similar to 
the data, but the calculated $2^+_2$ levels 
vary more rapidly with $N$ for the region $N \geqslant 68$ 
than the experimental values.

%
%
\begin{figure*}
\begin{center}
\includegraphics[width=.7\linewidth]{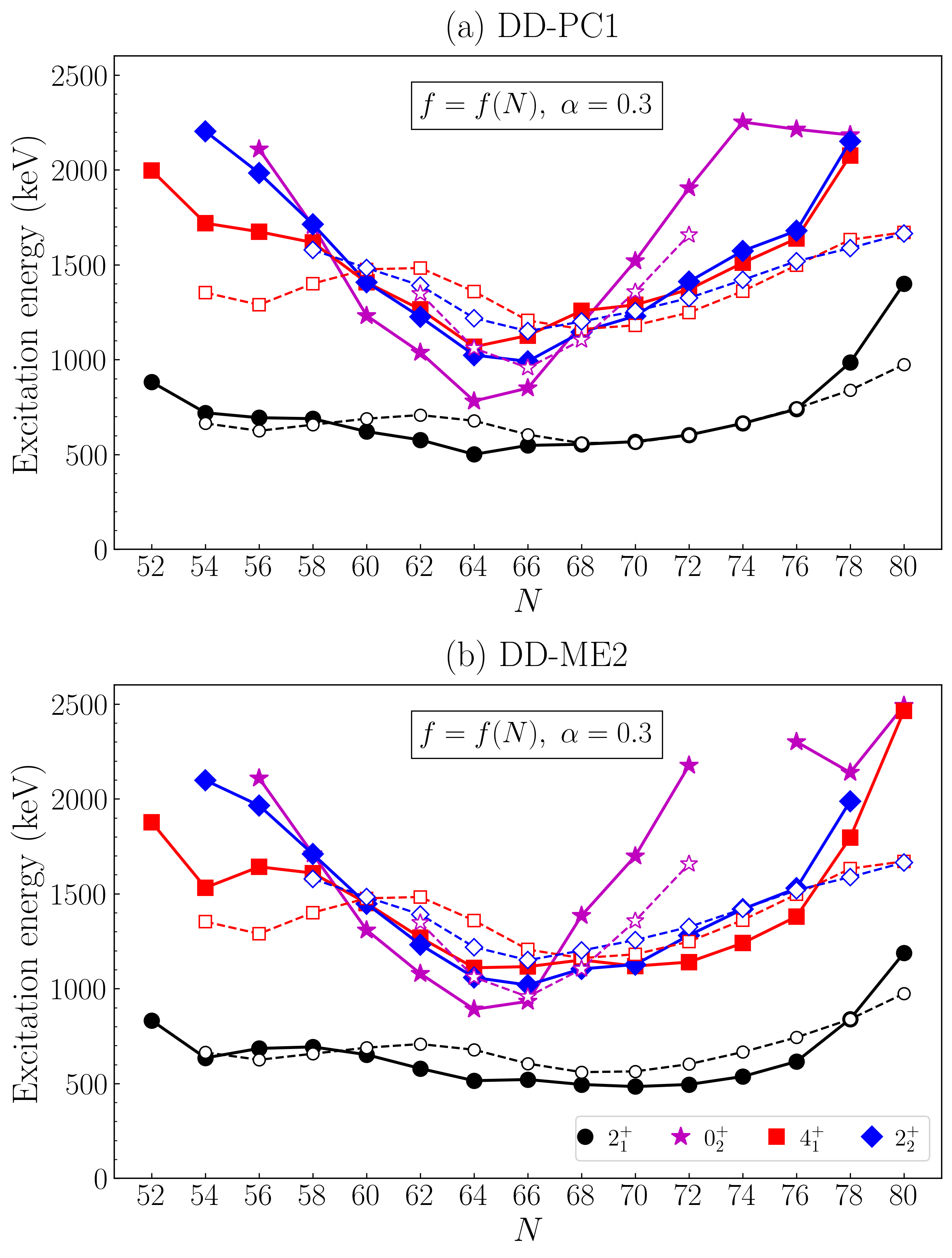}
\caption{
Predicted low-energy spectra 
by the RHB+QCH model based on the 
(a) DD-PC1 and (b) DD-ME2 EDFs, represented by 
the solid symbols connected by solid lines. 
The corresponding experimental data \cite{data} 
are shown as the open circles connected by 
broken lines.  
For both RHB-SCMF calculations, the $N$-dependent 
pairing strength $f(N)$ (\ref{eq:pair3}) is employed, 
and the MOI is scaled by 30 \% ($\alpha=0.3$). 
}
\label{fig:pc-me}
\end{center}
\end{figure*}

Figure~\ref{fig:pc-me}(a) 
summarizes the RHB-QCH results 
for the $2^+_1$, $4^+_1$, $0^+_2$, and $2^+_2$ states 
obtained from the DD-PC1 EDF employing the $N$-dependent 
pairing strength. 
The QCH results in the case of the DD-ME2 EDF 
are also shown in Fig.~\ref{fig:pc-me}(b). 
The results obtained from the DD-ME2 EDF are qualitatively 
similar to those from the DD-PC1 EDF, a major difference 
being perhaps the location of the $0^+_2$ energy level 
for the mid-shell nuclei with $62 \leqslant N \leqslant 66$. 
The RHB+QCH model employing the DD-ME2 EDF gives slightly 
higher lying $0^+_2$ states than in the case of the DD-PC1 EDF. 
In addition, the DD-ME2 result exhibits 
slightly lower-lying $2^+_1$, $4^+_1$, and $2^+_2$ 
levels on the neutron-deficient side 
than in the calculation with the DD-PC1 EDF. 
The results shown in Figs.~\ref{fig:pc-me}(a) 
and \ref{fig:pc-me}(b) illustrate the difficulty in reproducing 
the $0^+_2$ excitation energies in the mid-shell Te isotopes, 
irrespective of which of the two representative relativistic EDFs 
is used for the RHB-SCMF calculation. 
We have also confirmed that increasing the 
pairing strength and rotational MOI does not appear 
to improve the description of the $0^+_2$ levels 
in the case where the DD-ME2 functional is employed. 
The deficiency in reproducing the $0^+_2$ energies 
in the mid-shell region with the QCH framework 
could be mainly attributed to the topology 
of their PESs, which exhibit two closely lying 
oblate global and prolate local minima. 
The topology of the PESs, 
in turn, depends much on the properties of 
the underlying EDF in general, which is usually not specifically 
adjusted to particular nuclei under consideration.

%
%
\begin{figure}
\begin{center}
\includegraphics[width=\linewidth]{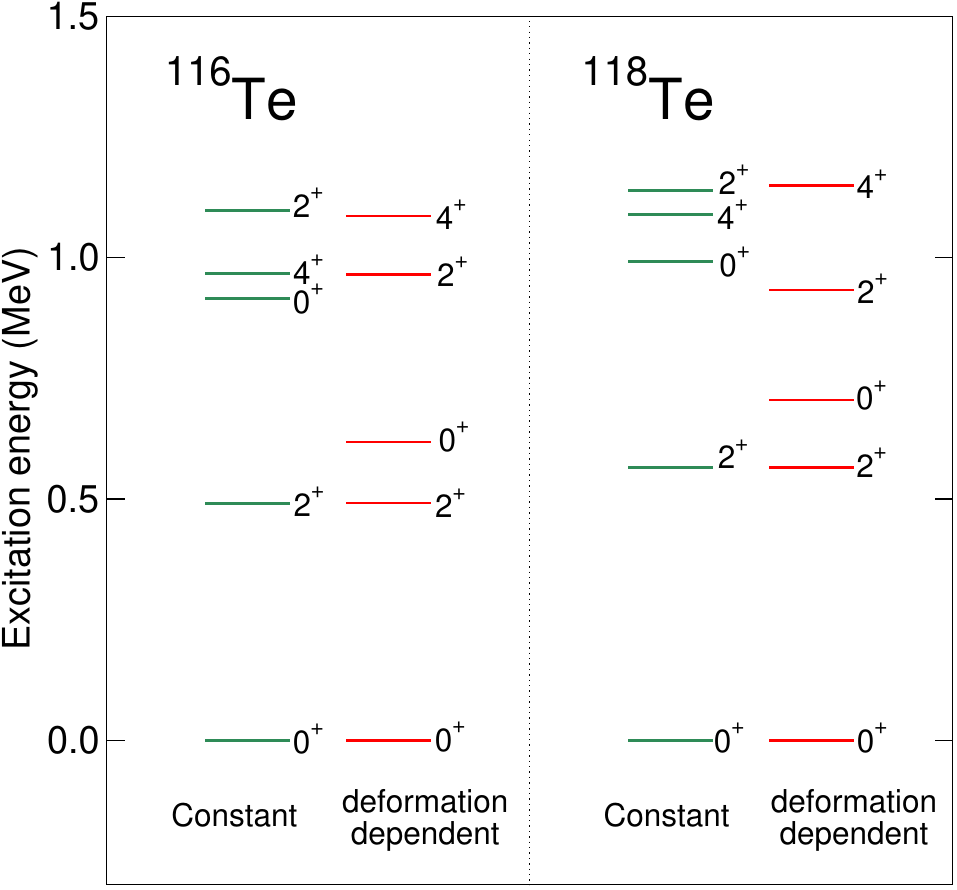}
\caption{
Low-energy spectra for $^{116}$Te and $^{118}$Te 
obtained with the constant 
and deformation-dependent 
collective mass parameters. 
The MOI in the latter calculation is here not 
scaled (i.e., $\alpha=0$), 
and the $2^+_1$ energy in the former 
calculation is adjusted to that in the latter.
The default pairing 
strength is employed in both calculations. 
}
\label{fig:mass}
\end{center}
\end{figure}

We now try to identify which ingredients of the 
collective Hamiltonian make essential contributions 
to determine the low-energy level structure, especially, the 
location of the $0^+_2$ energy level in the 
mid-shell Te isotopes. 
Here we specifically analyze the roles 
of the collective mass parameters, since they depend 
both on the $\beta$ and $\gamma$ deformations 
and are therefore influenced by the topology of the PES. 
In a recent QCH study using a Skyrme EDF 
\cite{washiyama2023}, the QCH energy spectra  
calculated with the deformation-dependent 
rotational MOIs and vibrational mass 
parameters were compared with those obtained with constant 
mass parameters, and it was suggested that 
the deformation-dependent 
mass parameters play an important role 
in accurately predicting the 
properties of the excited $0^+$ states in 
the $N\approx 28$ nuclei. 
Here we apply the above procedure to the studied 
Te nuclei, and perform additional RHB+QCH 
calculations with an assumption that 
all the mass parameters defined 
in Eqs.~(\ref{eq:vib}) and (\ref{eq:rot}) 
should be constant and equal to each other, i.e., 
$B_1=B_2=B_3=B_{\beta\beta}=B_{\gamma\gamma}\equiv B_0$, 
and that $B_{\beta\gamma}=0$, as considered 
in Ref.~\cite{washiyama2023}.

Figure~\ref{fig:mass} compares between 
the RHB+QCH energy spectra for $^{116}$Te 
and $^{118}$Te, obtained 
with the constant and deformation-dependent mass parameters 
for the QCH.  
Note that, for the latter calculation, the original MOI 
($\alpha=0$) is used, and that the parameter $B_0$ 
in the former calculation is determined to fit the 
$2^+_1$ energy obtained from the former. 
In both calculations, the default pairing 
strength ($f=1.0$) is employed. 
One can clearly see that for both $^{116,118}$Te 
the $0^+_2$ level obtained with the QCH employing the 
deformation-dependent collective mass parameters 
is substantially lower and closer 
in energy to the $2^+_1$ one 
than that resulting from the QCH employing the 
constant mass parameters. 
This finding suggests that the deformation-dependent 
collective masses, which are influenced by the $\beta$ as well as 
$\gamma$ softness of the PES, make significant contributions 
to describe location of the excited $0^+_2$ states 
relative to other states. 
A more thorough investigation on the 
effects of the mass parameters in a wider mass 
region within the relativistic EDF framework 
is an interesting future study.

Furthermore, a possible solution to improve description 
of the excited $0^+_2$ levels would be to 
introduce additional collective coordinates 
to the triaxial quadrupole deformations, such 
as the dynamical pairing degree of freedom. 
The collective Hamiltonian 
that includes quadrupole deformations and 
dynamical pairing correlations 
was shown to provide a more accurate description 
of the excited $0^+$ states and the bands built 
on them in the neighboring Xe isotopes and rare-earth 
nuclei with $N\approx 90$ \cite{xiang2020,nomura2020pv,xiang2024}.

As an alternative EDF-based 
approach to low-energy collective states, 
one could adopt the mapped IBM that 
explicitly takes into account configuration mixing 
between the normal and intruder states 
\cite{nomura2012sc,nomura2016sc}. 
Within this formalism, 
two independent IBM Hamiltonians, 
corresponding to 
the normal ($0p-0h$) and intruder ($2p-2h$) 
configurations, are determined by mapping 
the SCMF-PES onto the bosonic one, and are 
allowed to mix via certain mixing interactions 
(see Refs.~\cite{nomura2012sc,nomura2016sc} for details). 
This procedure has been applied to a number of 
mass regions in which shape coexistence is 
suggested to occur 
\cite{nomura2012sc,nomura2013hg,nomura2016sc,nomura2016zr}, 
and would shed lights on the quality of the 
relativistic EDFs and the low-lying $0^+$ states 
of the Te nuclei.

%
%
\begin{figure}
\begin{center}
\includegraphics[width=\linewidth]{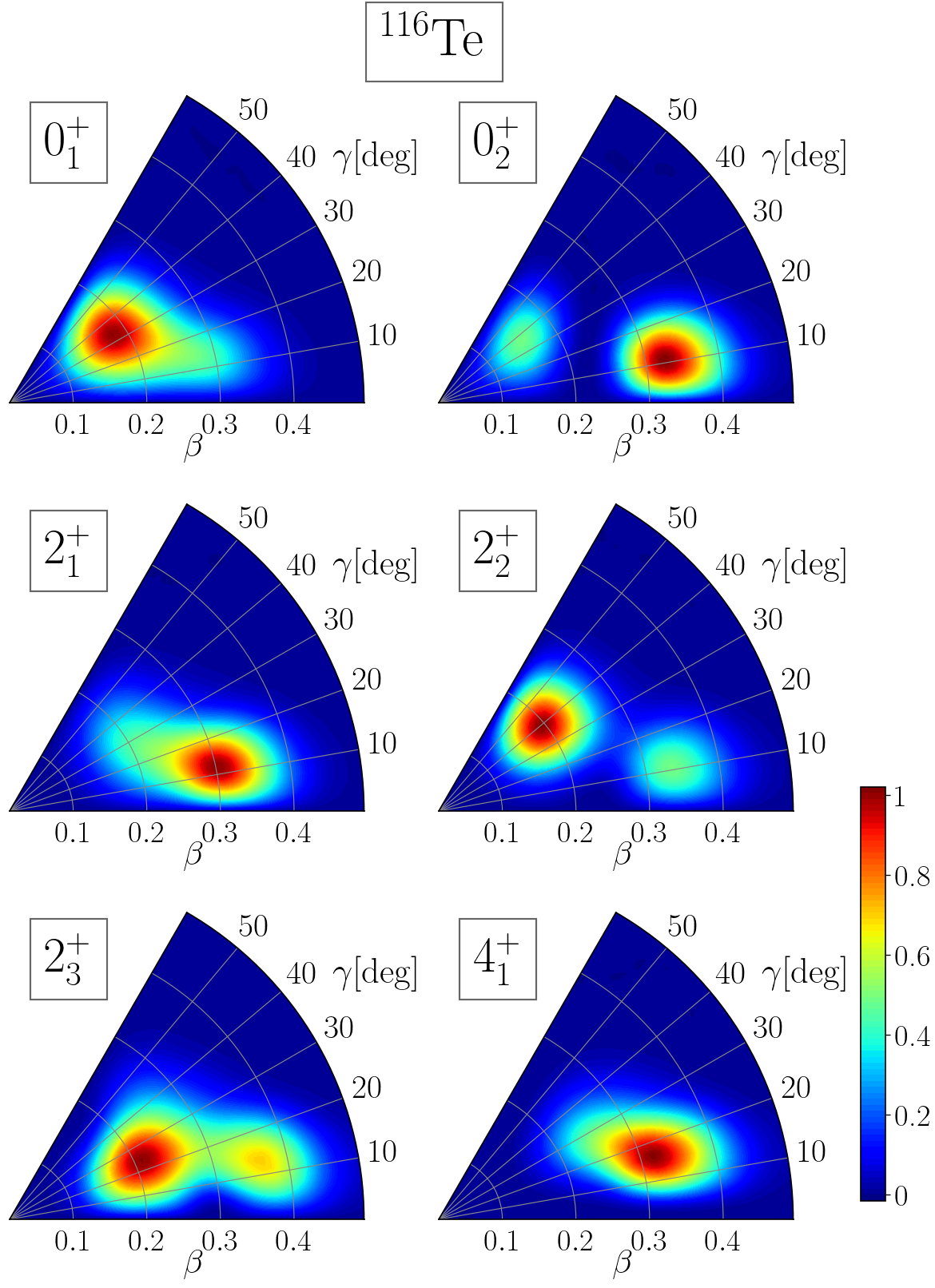}
\caption{Distributions of collective wave functions 
for the $0^+_{1}$, $0^+_2$, $2^+_{1}$, $2^+_{2}$, $2^+_{3}$, 
and $4^+_1$ states in the $(\beta,\gamma)$ plane 
of $^{116}$Te, 
which are obtained from the RHB+QCH 
method using the default 
pairing strength $V$ and MOI.}
\label{fig:cwf-te116}
\end{center}
\end{figure}

%
%
\begin{figure}
\begin{center}
\includegraphics[width=\linewidth]{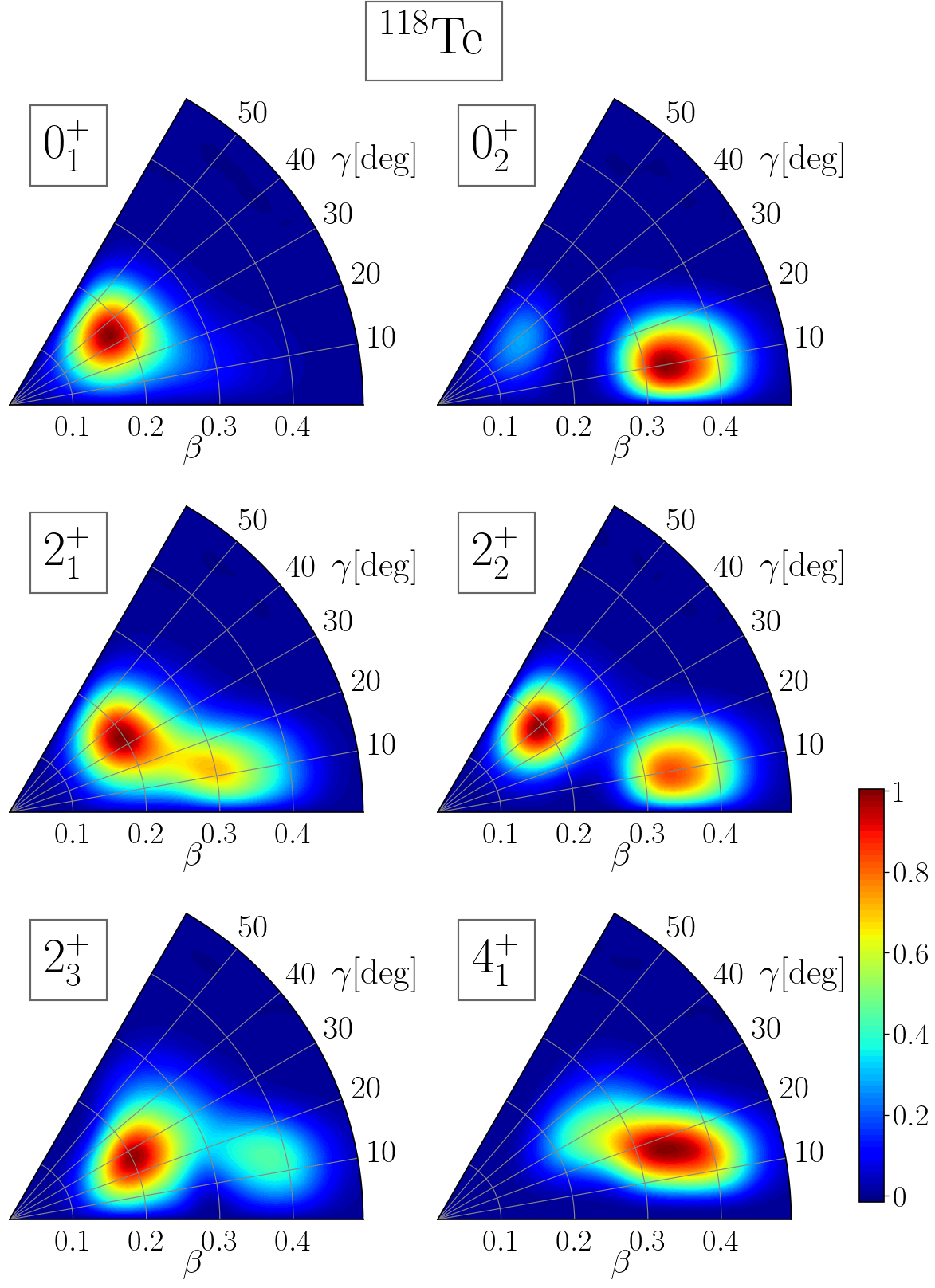}
\caption{Same as for Fig.~\ref{fig:cwf-te116}, but 
for $^{118}$Te.}
\label{fig:cwf-te118}
\end{center}
\end{figure}

\subsection{Collective wave functions}

The nature of a given calculated state can be 
analyzed in terms of the collective wave function 
obtained as the eigenvector of the QCH at 
each deformation. 
Figures~\ref{fig:cwf-te116} and \ref{fig:cwf-te118} 
show, respectively, distributions of the 
calculated collective wave functions for the 
$0^+_{1}$, $0^+_2$, $2^+_{1}$, $2^+_2$, $2^+_3$, and $4^+_1$ 
states for the $^{116}$Te and $^{118}$Te nuclei. 
Here only the results obtained by using 
the default pairing 
strength with $f=1.0$ in the RHB-SCMF calculation 
based on the DD-PC1 EDF and the 
default MOI with $\alpha=0$ are considered 
without a loss of generality. 
Note that qualitatively very similar results 
are obtained from the calculations with the 
modified pairing strengths and/or MOI, 
and from those using the DD-ME2 EDF. 
For both $^{116}$Te and $^{118}$Te 
nuclei, the wave function for the 
$0^+_1$ ground state is peaked at $\gamma\approx 36^{\circ}$, 
which corresponds to the oblate 
absolute minimum on the PES (see Fig.~\ref{fig:pes}). 
The $0^+_1$ wave function for the $^{116}$Te is, 
however, more extended toward the prolate side 
within the range $0.2 \leqslant \beta \leqslant 0.3$, 
than that of $^{118}$Te, and this implies 
strong shape mixing in the former nucleus. 
The $0^+_2$ wave functions for both nuclei 
exhibit a distinct peak near the prolate 
axis, that is, $\gamma \approx 12^{\circ}$ 
and $\beta \approx 0.32$, which appears to 
be associated with the local minimum near the 
prolate axis in the PES. 
The fact that the distinct peaks appear, 
one on the oblate side in the $0^+_1$ wave function 
and the other on the prolate side 
with larger $\beta$ deformation 
in the $0^+_2$ one, is in accordance with  
the presence of the closely lying 
oblate and prolate mean-field minima obtained in the 
RHB-SCMF calculation. 
From their collective wave functions, 
one could see that the ground state $0^+_1$ 
can be here attributed to a near oblate 
triaxial deformation with $\beta$ being 
$\approx 0.17$, while the $0^+_2$ state 
is constructed based on the configurations 
representing a near prolate triaxial 
deformation with larger $\beta$ ($>0.3$) value.

As one can see in Figs.~\ref{fig:cwf-te116} 
and \ref{fig:cwf-te118}, 
the nature of the $2^+_1$ 
collective wave function for the $^{116}$Te 
nucleus is rather different from that of the 
$^{118}$Te one: for the former 
the peak is observed near the prolate axis with 
$(\beta,\gamma)$ $\approx$ $(0.3, 12^{\circ})$, 
while for the latter a major peak is seen rather 
in the triaxial region,
$(\beta,\gamma)$ $\approx$ $(0.2, 36^{\circ})$, 
and a minor peak is also obtained 
near the prolate axis, 
$(\beta,\gamma)$ $\approx$ $(0.3, 10^{\circ})$. 
The $2^+_1$ as well as $0^+_2$ states of $^{116}$Te 
are, therefore, 
coming from the near prolate configurations that 
are supposed to originate from the intruder states. 
Concerning $^{118}$Te, 
since the $0^+_1$ and $2^+_1$ wave functions are 
similar to each other in that both have a major peak 
in the triaxial oblate region, these states 
are supposed to be mostly composed of the 
weakly oblate deformed 
states, associated with the global oblate minimum 
in the corresponding PESs.

There are two peaks visible in 
the distributions of the $2^+_2$ state 
for both Te nuclei, the major one in the 
triaxial oblate region and the other 
near the prolate axis. In addition, 
the peak on the prolate side in the $2^+_2$ 
wave function distribution for $^{118}$Te is higher 
than that of $^{116}$Te, 
meaning that both triaxial oblate and prolate 
components are more strongly mixed in the 
$2^+_2$ wave function for $^{118}$Te
than in the $^{116}$Te counterpart.

The $2^+_3$ collective wave functions 
for both $^{116}$Te and $^{118}$Te 
exhibit a major peak in the triaxial 
region with $\beta\approx 0.2$ and 
$\gamma \approx 25^{\circ}$, 
and a minor one near the prolate axis with 
$\beta \approx 0.37$ and $\gamma \approx 15^{\circ}$. 
These structures seem to be more or less similar 
to those found in the $2^+_2$ wave function 
distributions, which also exhibit 
two peaks on the oblate and prolate sides 
with slightly different ($\beta,\gamma$) coordinates. 
Given the distributions 
of the $2^+_2$ and $2^+_3$ collective wave functions, 
we find certain similarities 
between the structures of these two states.

For both $^{116}$Te and $^{118}$Te, the 
distribution of the calculated $4^+_1$ 
collective wave function essentially has a 
single peak at ($\beta,\gamma$) 
$\approx$ ($0.3 - 0.35, 15^{\circ}$), and this 
suggests that the $4^+_1$ 
state is mainly composed of the near prolate, 
triaxial configuration corresponding 
to the prolate local minimum on the 
mean-field PES. 
For the $^{116}$Te nucleus, the $0^+_2$, $2^+_1$, 
and $4^+_1$ collective wave functions exhibit a distinct 
peak on the prolate side, and this implies 
that these three states are, to a large extent, 
accounted for by the prolate deformed 
configurations. 

%
%
\begin{figure*}[ht]
\begin{center}
\includegraphics[width=.49\linewidth]{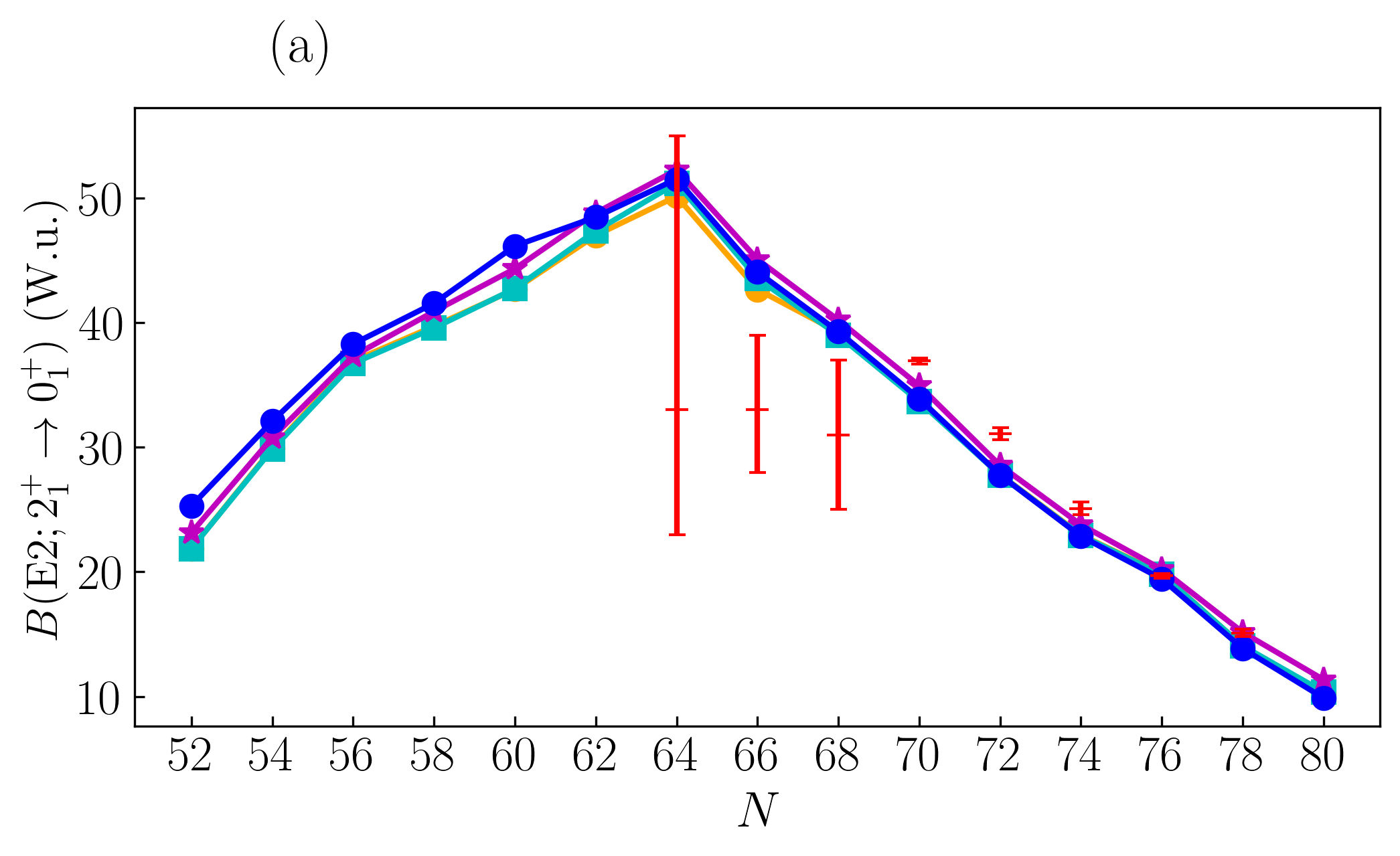}
\includegraphics[width=.49\linewidth]{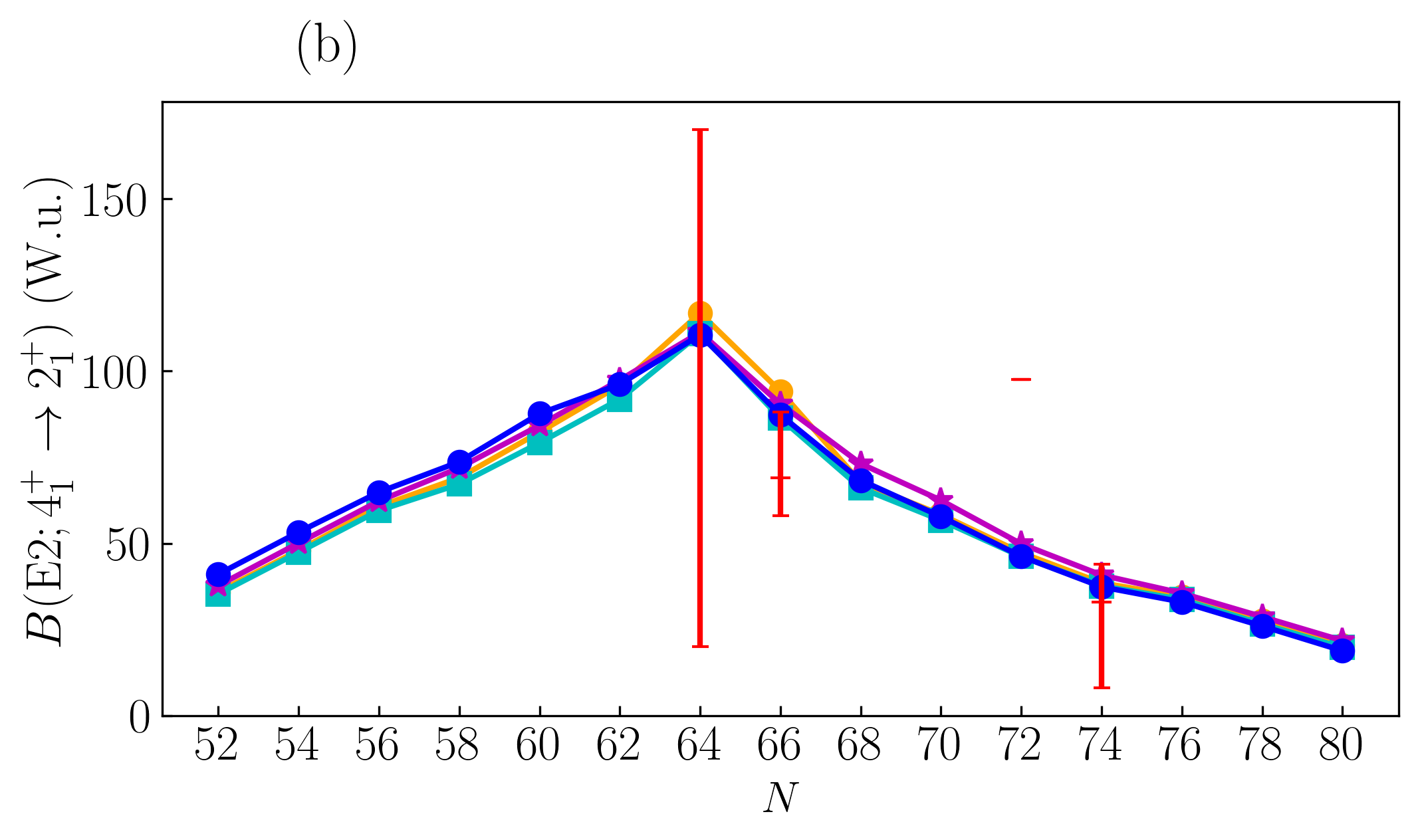}\\
\includegraphics[width=.49\linewidth]{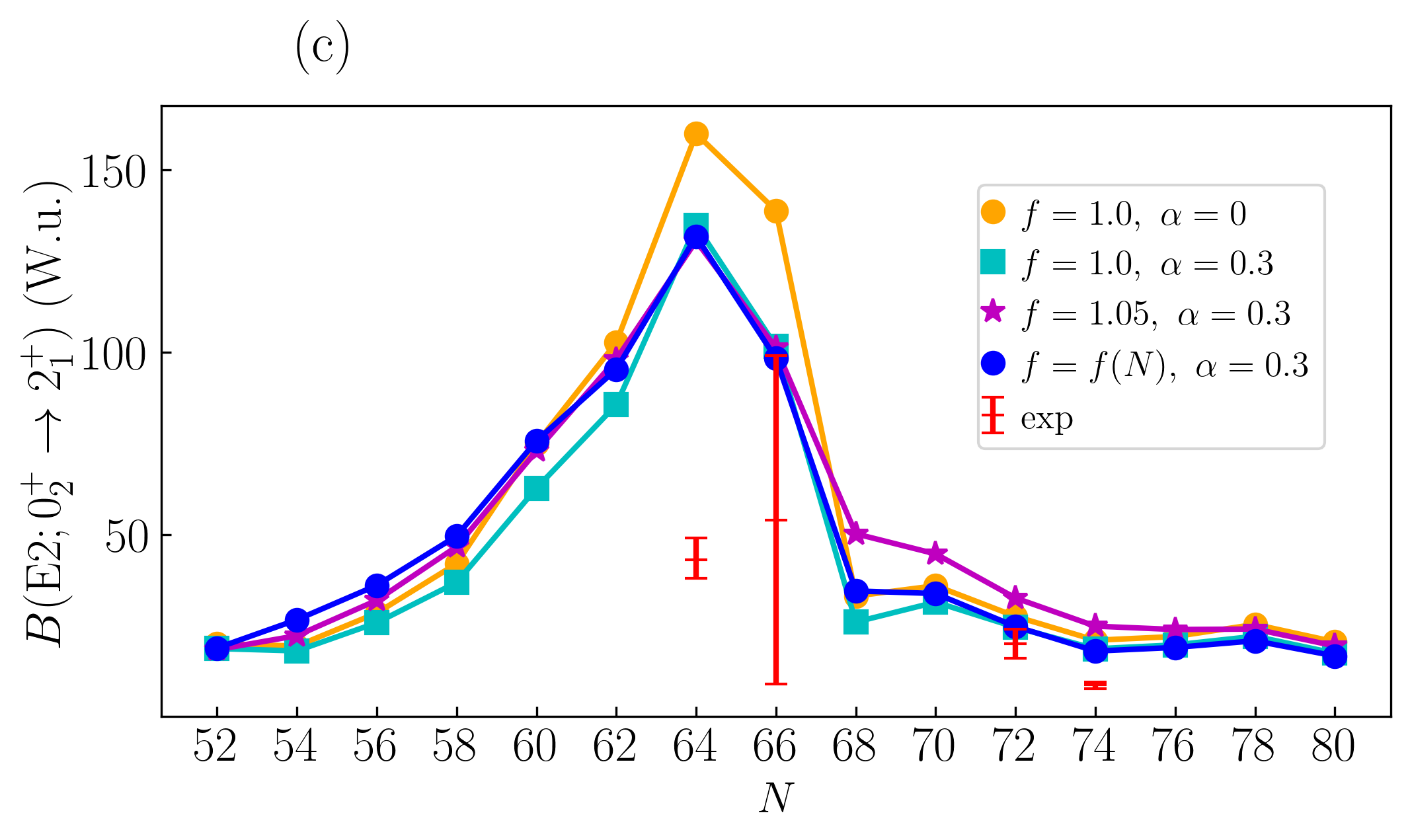}
\includegraphics[width=.49\linewidth]{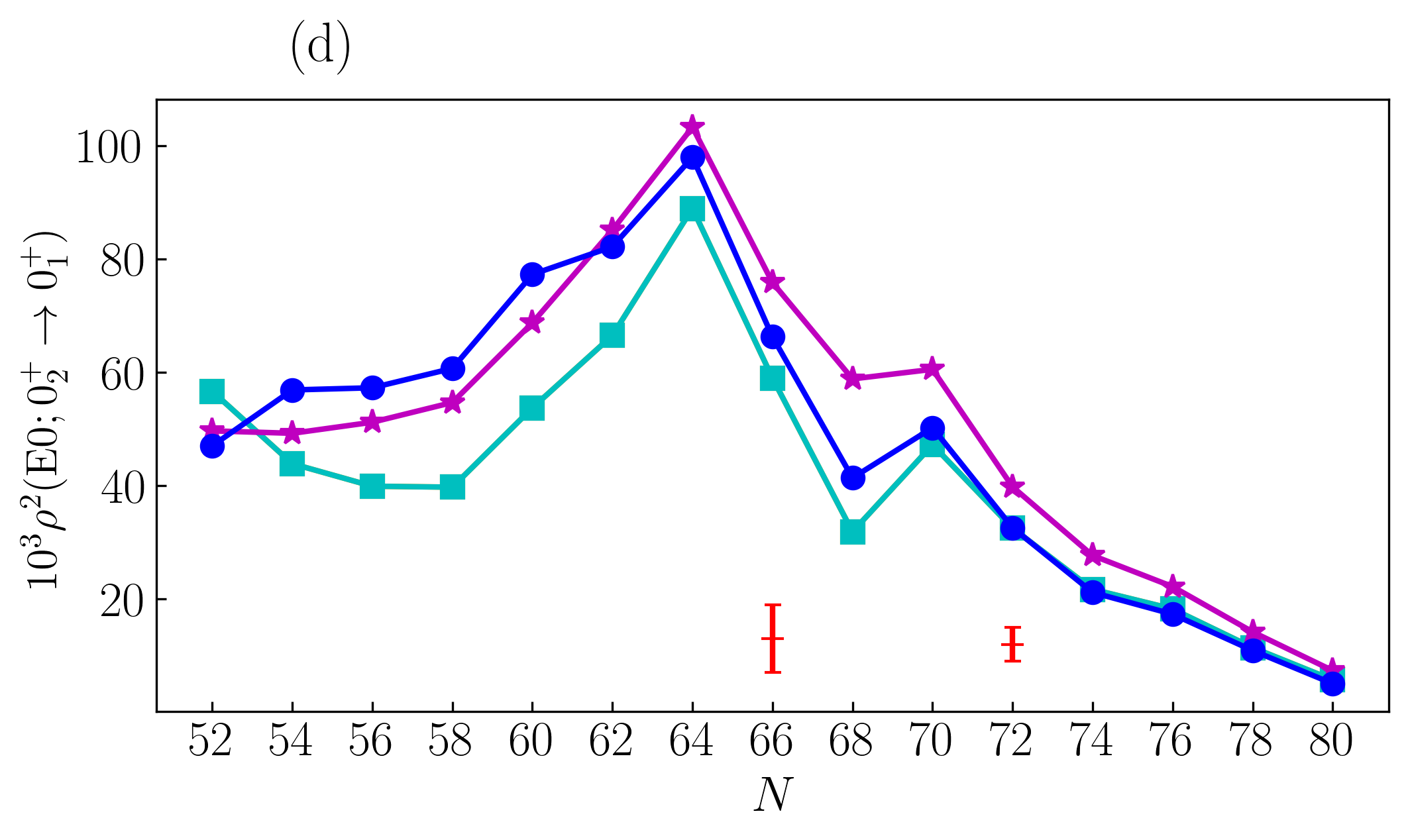}\\
\caption{
(a) $B(E2;2^+_1 \to 0^+_1)$, 
(b) $B(E2;4^+_1 \to 2^+_1)$, and 
(c) $B(E2;0^+_2 \to 2^+_1)$ values 
in Weisskopf units (W.u.), 
and (d) $\rho^2(E0;0^+_2 \to 0^+_1)\times 10^3$ value 
for the $^{104-132}$Te isotopes as functions 
of $N$, computed 
within the RHB+QCH method with 
the factor $f$ for the pairing strength 
and the scaling factor $\alpha$ for the MOI 
equal to 
(i) $f=1.0, \alpha=0$, 
(ii) $f=1.0, \alpha=0.3$, 
(iii) $f=1.05, \alpha=0.3$, and 
(iv) $f=f(N), \alpha=0.3$. 
The DD-PC1 EDF is used. 
Experimental $B(E2)$ values are taken from 
Ref.~\cite{vonspee2024} for $^{116}$Te, 
from Ref.~\cite{mihai2011} for $^{118}$Te, 
and from Ref.~\cite{data} for all the other nuclei. 
The experimental $\rho^2(E0;0^+_2 \to 0^+_1)\times 10^3$ 
values for $^{118}$Te and $^{124}$Te 
are taken from Ref.~\cite{kibedi2005}. 
}
\label{fig:e2}
\end{center}
\end{figure*}

\subsection{Electric quadrupole and quadrupole properties}

Figure~\ref{fig:e2} exhibits the calculated 
$B(E2;2^+_1 \to 0^+_1)$, 
$B(E2;4^+_1 \to 2^+_1)$, and 
$B(E2;0^+_2 \to 2^+_1)$ values in Weisskopf units (W.u.), 
and $\rho^2(E0;0^+_2 \to 0^+_1)\times 10^3$ values 
for the $^{104-132}$Te isotopes as functions of the 
neutron number $N$. 
The four different RHB+QCH results 
with 
(i) $f=1.0$ and $\alpha=0$, 
(ii) $f=1.0$ and $\alpha=0.3$, 
(iii) $f=1.05$ and $\alpha=0.3$, and 
(iv) $N$-dependent $f$ (\ref{eq:pair3}) and $\alpha=0.3$
are compared, similarly to Fig.~\ref{fig:level}. 
Experimental data for the $B(E2)$ transition rates 
are available for $^{116}$Te \cite{vonspee2024}, 
$^{118}$Te \cite{mihai2011}, 
$^{120}$Te, and heavier Te nuclei \cite{garrett2022,data}.

The predicted $B(E2;2^+_1 \to 0^+_1)$, 
and $B(E2;4^+_1 \to 2^+_1)$ values both monotonically  
increase toward the middle of the neutron major shell, 
$N=64$ or 66, reflecting the growing 
quadrupole collectivity in the open shell nuclei. 
The RHB+QCH calculation gives results that are 
larger than the measured 
$B(E2;2^+_1 \to 0^+_1)$ values at $N=66$ and 68, 
while the agreement with experiment looks reasonable 
for $N\geqslant 70$.
Concerning $^{116}$Te, the present calculation 
provides the $B(E2;2^+_1 \to 0^+_1)$ values of 
$\approx 50$ W.u., which are within the error bar of the 
measured one \cite{vonspee2024}, $33^{+22}_{-10}$ W.u.
The $B(E2;4^+_1 \to 2^+_1)$ value for the same 
nucleus is here calculated to be $\approx 115$ W.u., 
while the recent measurement suggested the lower 
and upper limits of this transition rate to be 
20 and 170 W.u., respectively. 
The phenomenological IBM-1 calculation assuming 
the single U(5) configuration, which was 
also reported in Ref.~\cite{vonspee2024}, 
provided the 
$B(E2;2^+_1 \to 0^+_1)$ and  
$B(E2;4^+_1 \to 2^+_1)$ values that are 
about half the ones obtained in the 
present calculation. 
All the four RHB+QCH calculations, shown in 
Figs.~\ref{fig:e2}(a) and \ref{fig:e2}(b), 
appear to give essentially identical results.

The $B(E2;0^+_2 \to 2^+_1)$ transition rate 
is considered an indicator of the shape mixing 
and coexistence. 
As seen from Fig.~\ref{fig:e2}(c), 
the calculated values of this quantity near the 
mid-shell nuclei with $N=60-66$ 
are generally large, and are almost in the 
same order of magnitude as the 
$B(E2;4^+_1 \to 2^+_1)$ transition rates. 
This systematic corroborates the 
fact that substantial degree of 
softness along the triaxial deformation 
is present in the mean-field PESs. 
The behavior of the predicted $B(E2;4^+_1 \to 2^+_1)$ 
is further in accordance with the result that 
the $0^+_2$ energy levels for these nuclei 
are calculated to be especially low. 
The RHB+QCH calculation suggests for 
the $^{116}$Te nucleus 
the $0^+_2 \to 2^+_1$ $E2$ transition rate of 160 W.u., 
when the original pairing ($f=1.0$) and 
MOI ($\alpha=0$) are employed. 
An effect of increasing the MOI is such that 
the $B(E2;0^+_2\to 2^+_1)$ values are lowered 
for the $N=64$ and 66 isotopes. 
With the increased MOI, the predicted 
$B(E2;0^+_2\to 2^+_1)$ values become lower, 
being closer to the upper limit of the 
measured value. 
Modification to the pairing strength, however, 
does not alter the $B(E2;0^+_2\to 2^+_1)$ 
transition rates. 
The corresponding experimental value for $^{116}$Te 
available from \cite{vonspee2024} is $43^{+6}_{-5}$ W.u., 
which is much lower than the calculated 
values by a factor of 3 to 4. 
The enhanced $B(E2;0^+_2\to 2^+_1)$ 
transition rate illustrates that the present 
RHB+QCH calculation rather overestimates 
the degree of shape mixing for the near mid-shell 
Te nuclei, and this feature persists even 
if the increased pairing strength 
is employed in the RHB calculation.

The $E0$ transition between the $0^+_1$ 
and $0^+_2$ states is another signature of the 
shape mixing and evolution of collectivity. 
The predicted $\rho^2(E0;0^+_2 \to 0^+_1)\times 10^3$ 
values, shown in Fig.~\ref{fig:e2}, exhibit 
a monotonic increase toward the mid-shell, 
with the maximal value being found at $N=64$. 
This appears to be a similar 
systematic behavior as a function of $N$ 
to the $B(E2)$ values. 
Experimental information about the $E0$ transitions 
is scarce for Te isotopes, as only the available data 
are $\rho^2(E0;0^+_2 \to 0^+_1)\times 10^3=13 \pm 6$ 
and $12 \pm 3$ for $^{118}$Te and $^{124}$Te, 
respectively \cite{kibedi2005}. 
The present calculation significantly overestimates 
these values. 
The calculated 
$\rho^2(E0;0^+_2 \to 0^+_1)\times 10^3$ 
values for lighter Te near the neutron major 
shell $N=50$ are, however, also large, 
even though the quadrupole collectivity 
is expected to be much less significant 
for near magic nuclei than in the open-shell nuclei. 
The $\rho^2(E0;0^+_2 \to 0^+_1)\times 10^3$ values 
are here rather sensitive to the pairing strength, 
as it affects the energies and wave functions 
of the $0^+_2$ states. 
The increase of the MOI does not change this 
transition at all, hence the results with the 
default and increased MOIs are identical 
in the plot.

%
%
\begin{figure*}[ht]
\begin{center}
\includegraphics[width=.8\linewidth]{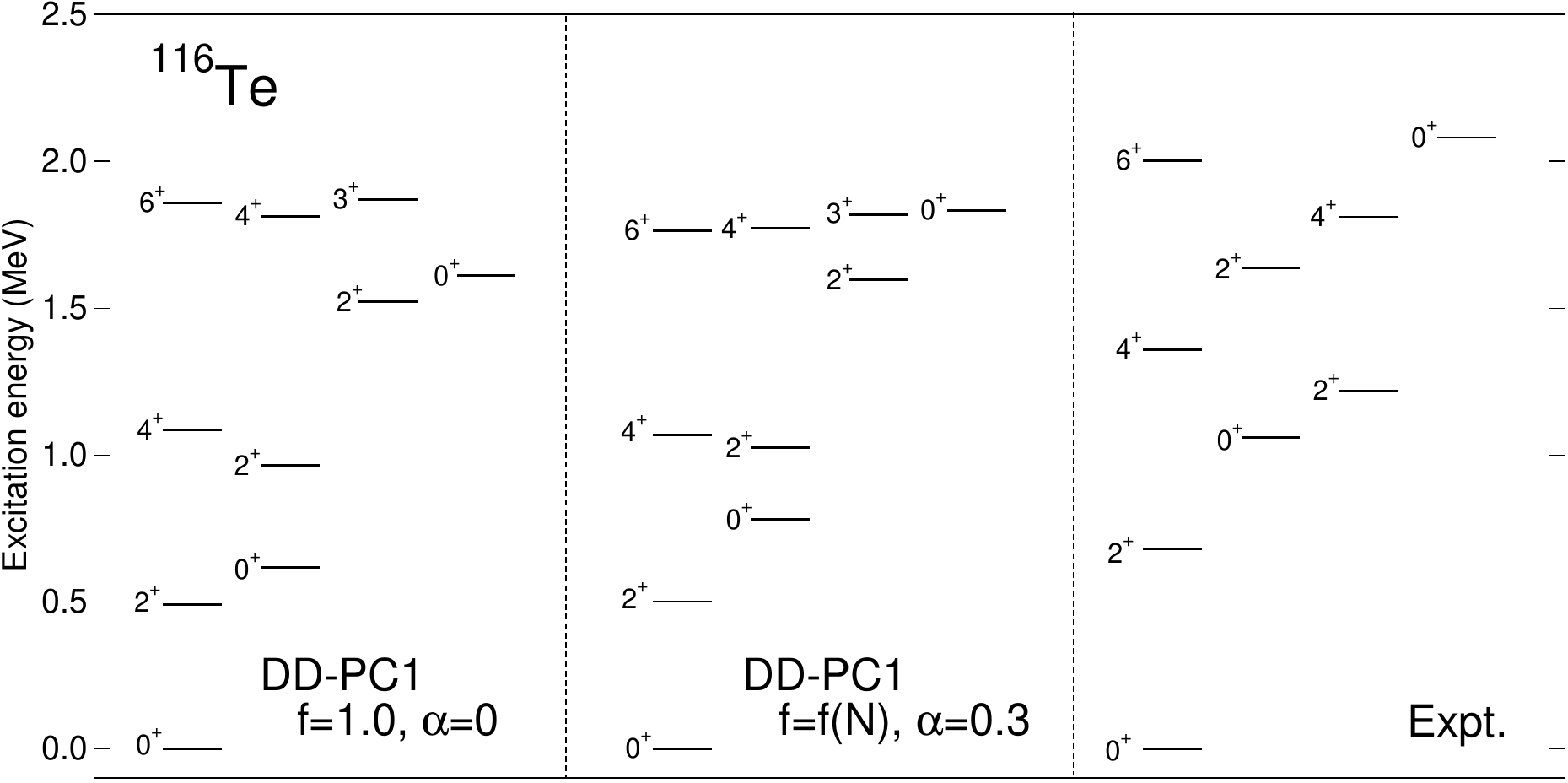}
\caption{
Low-energy level scheme for $^{116}$Te 
predicted by the RHB+QCH model using the DD-PC1 EDF 
and separable pairing force with 
(left) the default strength and the MOI, 
and (middle) the $N$-dependent pairing strength $f(N)$ 
and increased MOI with $\alpha=0.3$. 
On the right-hand side shown are the 
experimental data \cite{vonspee2024}. 
}
\label{fig:te116}
\end{center}
\end{figure*}

%
%
\begin{figure*}[ht]
\begin{center}
\includegraphics[width=.8\linewidth]{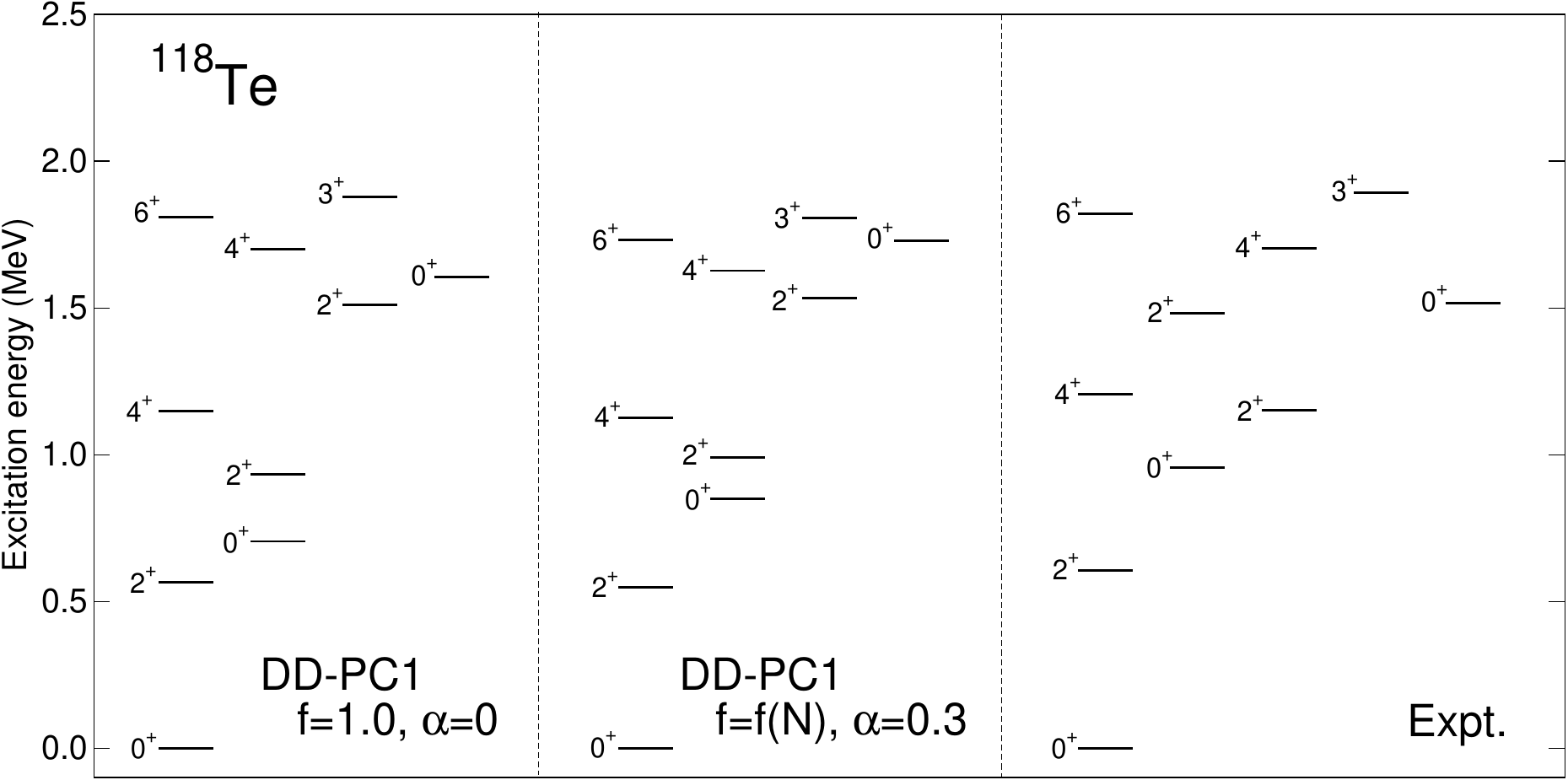}
\caption{
Same as for Fig.~\ref{fig:te116}, but for the $^{118}$Te 
nucleus and the 
experimental data are taken from Ref.~\cite{mihai2011}. 
}
\label{fig:te118}
\end{center}
\end{figure*}

\subsection{Level scheme for $^{116}$Te and $^{118}$Te}

We show in Figs.~\ref{fig:te116} and \ref{fig:te118} 
the predicted low-energy level schemes for $^{116}$Te 
and $^{118}$Te. 
The calculated results are compared with 
the experimental data available from Refs.~\cite{vonspee2024} 
and \cite{mihai2011} for $^{116}$Te and $^{118}$Te, 
respectively. 
Note that the experimental level at 1637 keV 
is assigned to be the $3^+$ state in the 
NNDC database \cite{data}, but in the new 
measurement of Ref.~\cite{vonspee2024} it 
has been identified as the $2^+_3$ state. 
The experimental level scheme shown on the 
right-hand side of Fig.~\ref{fig:te116} follows 
the latter assignment. 
The RHB+QCH results with the $N$-dependent 
pairing strength and increased MOI by 30 \% are 
compared with those with the original 
pairing strength ($f=1.0$) and MOI ($\alpha=0$), 
and with the experimental data.

For $^{116}$Te, both calculations provide 
similar results for the ground-state band, 
which look compressed as compared to the 
experimental one. 
The $0^+_2$ level obtained with the modified pairing 
and MOI is closer to the measured one than with 
the original pairing and MOI. 
The $0^+_2$ states is supposed to be the bandhead of the 
possible $K^\pi=0^+_2$ band connected by dominant 
in-band $\Delta I=2$ $E2$ transitions.
In Fig.~\ref{fig:te116}, 
the $2^+_2$ level in both calculations is 
depicted as belonging to 
the $0^+_2$ band, and the $2^+_3$ as 
the bandhead of a band resembling $\gamma$-vibrational 
band connected by the 
$\Delta I=1$ $E2$ transitions. 
As noted earlier, 
the $K^\pi=0^+$ and $K^\pi=2^+$ components 
are considerably mixed in the $2^+_2$ and $2^+_3$ states: 
41 \% and 59 \%
(56 \% and 44 \%) of the $2^+_2$ ($2^+_3$) 
wave function is accounted for by the 
$K^\pi=0^+$ and $2^+$ components, respectively, 
in the case of the modified pairing and MOI. 
Similar degrees of $K$ mixing in 
the $2^+_{2}$ and $2^+_3$ states are obtained 
from the calculation using the default pairing and MOI. 
The strong $K$ mixing between the $2^+_2$ and $2^+_3$ states 
is also evident from the fact that these states exhibit 
sizable $E2$ transitions to the $0^+_2$ state, 
with the corresponding $B(E2)$ rates being 
40 (35) W.u. and 23 (35) W.u., respectively, 
when the original pairing strength $f=1.0$ 
[$N$-dependent strength $f(N)$] is used. 
Recall also that the distributions 
of the $2^+_2$ and $2^+_3$ collective 
wave functions in the 
$(\beta,\gamma)$ plane (cf. Fig.~\ref{fig:cwf-te116}) 
both exhibit 
a peak at the weak deformation $\beta\approx 0.2$ 
in triaxial region, and the other peak with 
larger $\beta>0.3$ near the prolate axis. 
The $0^+_3$ level is reproduced rather well 
by the RHB+QCH calculation, especially that 
with the modified pairing strength.

As it is seen in Fig.~\ref{fig:te118}, 
basically similar level scheme to $^{116}$Te 
is predicted for $^{118}$Te. 
Overall, the calculated excitation energies 
for $^{118}$Te agree with the experimental data \cite{mihai2011} 
comparatively well. 
In particular, the predicted $0^+_2$ energy level with the 
$N$-dependent pairing and increased MOI is higher than 
and closer to experimental counterpart than 
that with the default pairing and MOI.

Table~\ref{tab:e2} summarizes the calculated 
$B(E2)$ transition rates and electric quadrupole 
moments $Q(2^+)$ of low-energy states 
in $^{116}$Te and $^{118}$Te. 
The experimental data are taken from 
Ref.~\cite{vonspee2024} for $^{116}$Te, 
and Ref.~\cite{mihai2011} for $^{118}$Te. 
The RHB+QCH calculation provides a substantially 
strong $B(E2;0^+_2 \to 2^+_1)$ rate especially for 
$^{116}$Te, which amounts to approximately 
by a factor of three larger than than 
the corresponding data. 
For the other $E2$ transitions of the 
non-yrast states, the QCH 
generally gives large $B(E2)$ rates, whereas 
the experimental information is rather limited 
for $^{116}$Te as only the lower limits are 
known. 
More experimental $B(E2)$ data are available 
for the $^{118}$Te nucleus. 
The RHB+QCH in most cases gives larger $B(E2)$'s 
than the experimental data, in particular, 
for the $6^+_1 \to 4^+_1$ and $0^+_2 \to 2^+_1$ 
$E2$ transitions. 
The calculation with the $N$-dependent 
pairing strength with $f(N)$ and increased 
MOI gives the $B(E2;0^+_2 \to 2^+_1)$ transition 
probability that is by 30 \% lower than 
that obtained with the default pairing 
and MOI. The reduction is due to the change 
in the MOI. 
The experimental $B(E2;6^+_1 \to 4^+_1)$ value 
that is in the same order of magnitude 
as the $B(E2;4^+_1 \to 2^+_1)$ value 
seems to be an effect of configuration mixing. 
The present calculation is not able to 
account for this systematic.

Experimental information about the $Q(2^+)$ 
moments is not available for both nuclei, 
but the predicted values with the default (modified) 
pairing force, 
$Q(2^+_1)=-0.75$ ($-0.62$) $e$b
and $-0.42$ ($-0.25$) $e$b for $^{116}$Te and 
$^{118}$Te, respectively, are similar 
to the observed ones, both in magnitude and sign, 
for the heavier Te isotopes \cite{stone2005}. 

%
\begin{table}
\caption{\label{tab:e2}
Calculated $B(E2)$ rates (in W.u.) and electric 
quadrupole moments $Q(2^+_{1,2})$ (in $e$b) for the $^{116}$Te 
and $^{118}$Te nuclei, within the RHB+QCH 
calculations with the 
default pairing $f=1.0$ and MOI $\alpha=0$ (column 4, 
denoted as ``default''), 
and with the $N$-dependent pairing strength $f=f(N)$ 
and increased MOI with $\alpha=0.3$ (column 5, 
denoted as ``modified''). 
The DD-PC1 EDF is employed. 
Experimental data for 
$^{116}$Te and $^{118}$Te are adopted from 
Refs.~\cite{vonspee2024} and \cite{mihai2011}, 
respectively. 
}
 \begin{center}
 \begin{ruledtabular}
  \begin{tabular}{lcccc}
 & & \multirow{2}{*}{Expt.} & \multicolumn{2}{c}{RHB+QCH} \\
\cline{4-5}
 & & & default & modified \\
\hline
$^{116}$Te
& $B(E2;2^+_{1} \to 0^+_{1})$ & $33^{+22}_{-10}$ & 50 & 51 \\
& $B(E2;2^+_{2} \to 0^+_{1})$ & & 2.5 & 1.1 \\
& $B(E2;2^+_{2} \to 0^+_{2})$ & & 40 & 35 \\
& $B(E2;2^+_{2} \to 2^+_{1})$ & $41^{+6}_{-5}$ & 55 & 69 \\
& $B(E2;2^+_{3} \to 0^+_{1})$ & & 0.10 & 0.088 \\
& $B(E2;2^+_{3} \to 0^+_{2})$ & $>20$ & 23 & 35 \\
& $B(E2;2^+_{3} \to 2^+_{1})$ & $>3$ & 5.5 & 3.7 \\
& $B(E2;2^+_{3} \to 2^+_{2})$ & & 36 & 32 \\
& $B(E2;3^+_{1} \to 2^+_{3})$ & & 63 & 63 \\
& $B(E2;4^+_{1} \to 2^+_{1})$ & $>20$ and $<170$ & 117 & 110 \\
& $B(E2;4^+_{2} \to 2^+_{1})$ & $>0.7$ & 1.5 & 2.4 \\
& $B(E2;4^+_{2} \to 2^+_{2})$ & $>5$ & 69 & 68 \\
& $B(E2;4^+_{2} \to 4^+_{1})$ & & 24 & 35 \\
& $B(E2;6^+_{1} \to 4^+_{1})$ & $>27$ & 166 & 161 \\
& $B(E2;0^+_{2} \to 2^+_{1})$ & $43^{+6}_{-5}$ & 160 & 132 \\
& $B(E2;0^+_{3} \to 2^+_{1})$ & & 0.016 & 0.33 \\
& $B(E2;0^+_{3} \to 2^+_{2})$ & & 65 & 68 \\
& $B(E2;0^+_{3} \to 2^+_{3})$ & & 3.3 & 2.9 \\
& $Q(2^+_1)$ & & $-0.75$ & $-0.62$ \\
& $Q(2^+_2)$ & & $0.051$ & $0.050$ \\
[1.0ex]
$^{118}$Te
& $B(E2;2^+_{1} \to 0^+_{1})$ & $33^{+6}_{-5}$ & 43 & 44 \\
& $B(E2;2^+_{2} \to 0^+_{1})$ & $1.8^{+0.7}_{-0.4}$ & 1.2 & 0.11 \\
& $B(E2;2^+_{2} \to 0^+_{2})$ & & 62 & 56 \\
& $B(E2;2^+_{2} \to 2^+_{1})$ & $28^{+13}_{-7}$ & 72 & 77 \\
& $B(E2;2^+_{3} \to 0^+_{1})$ & & 0.025 & 0.011 \\
& $B(E2;2^+_{3} \to 0^+_{2})$ & $60^{+35}_{-17}$ & 22 & 36 \\
& $B(E2;2^+_{3} \to 2^+_{1})$ & $3.4^{+1.9}_{-0.9}$ & 13 & 10 \\
& $B(E2;2^+_{3} \to 2^+_{2})$ & $9.5^{+8.5}_{-3.5}$ & 28 & 28 \\
& $B(E2;3^+_{1} \to 2^+_{3})$ & & 61 & 59 \\
& $B(E2;4^+_{1} \to 2^+_{1})$ & $69^{+19}_{-11}$ & 94 & 87 \\
& $B(E2;4^+_{2} \to 2^+_{1})$ & $<1.1$ & 7.9 & 5.2 \\
& $B(E2;4^+_{2} \to 2^+_{2})$ & $<210$ & 63 & 70 \\
& $B(E2;4^+_{2} \to 4^+_{1})$ & $<160$ & 27 & 43 \\
& $B(E2;6^+_{1} \to 4^+_{1})$ & $80^{+14}_{-10}$ & 178 & 162 \\
& $B(E2;0^+_{2} \to 2^+_{1})$ & $54 \pm 45$ & 139 & 98 \\
& $B(E2;0^+_{3} \to 2^+_{1})$ & $1.3 \pm 45$ & 3.6 & 5.1 \\
& $B(E2;0^+_{3} \to 2^+_{2})$ & $100$ & 52 & 59 \\
& $Q(2^+_1)$ & & $-0.42$ & $-0.25$ \\
& $Q(2^+_2)$ & & $-0.37$ & $-0.40$ \\
 \end{tabular}
 \end{ruledtabular}
 \end{center}
\end{table}

\section{Summary\label{sec:summary}}

Evolution and coexistence of the intrinsic shape and the 
corresponding collective excitations in the 
even-even $^{104-132}$Te isotopes have been 
investigated using the theoretical framework of 
the relativistic EDF. 
The starting point of the present analysis 
was the constrained RHB-SCMF 
calculations that employ the DD-PC1 and DD-ME2 EDFs, 
and the separable pairing force of finite range. 
By using as a microscopic input 
the SCMF solutions, the deformation-dependent mass parameters 
and moments of inertia, and collective potential 
have been completely determined.

We found in the resultant quadrupole triaxial $(\beta,\gamma)$ 
PESs for the near mid-shell nuclei, $^{114-120}$Te, 
coexisting two minima that appear close in energy 
to each other: a weakly deformed oblate 
global minimum and a strongly deformed triaxial 
local minimum near the prolate ($\gamma=0^\circ$) axis. 
The calculated $0^+_2$ level has been shown 
to exhibit a parabolic 
behavior as a function of $N$, which is considered 
a signature of shape coexistence as expected 
empirically. 
The present RHB+QCH calculation, however, has yielded 
for the mid-shell Te nuclei significantly low-energy 
$0^+_2$ levels as compared to the 
experimental ones. 
The description of the $0^+_2$ 
excitation energies in the mid-shell region 
is supposed to be, to a large extent, sensitive to 
the topology of the $(\beta,\gamma)$ PESs, 
that is, the presence of the two closely lying 
oblate global and prolate local mean-field minima. 
These characteristics of the PESs, in turn, 
reflect the nature of the underlying EDF. 
The calculated $B(E2)$ and $\rho^2(E0)$ 
values also served as an indicator for the 
shape coexistence as well as growing quadrupole collectivity
toward the neutron mid-shell $N=66$. 
Here the $B(E2; 0^+_2 \to 2^+_1)$ and 
$\rho^2(E0; 0^+_2 \to 0^+_1)$ values for the 
mid-shell Te nuclei with $N\approx 64$ 
have been calculated to be 
too large, as compared with the available data, 
probably because a substantial degree of shape mixing 
occurs in the model calculation due to 
the softness in both $\beta$ and $\gamma$ 
deformations.

We have also studied the sensitivity of 
the spectroscopic results 
to the pairing strength employed 
in the RHB-SCMF calculation. 
It was shown that by increasing the pairing strength 
energy levels of the excited states were systematically raised, 
while the excitation energies for those nuclei with 
$N\geqslant 68$ were predicted to be 
even higher than the experimental ones as the 
$N=82$ major shell closure is approached. 
By using the $N$-dependent 
scaling factor for the pairing strength, 
certain improvements of the RHB+QCH description 
have been made, especially on the neutron-rich side.  
In addition, the deformation-dependent 
mass parameters in the collective Hamiltonian 
were shown to make an important contribution to 
determine the location of the $0^+_2$ energy level 
relative to the $2^+_1$ one 
of the near mid-shell nuclei $^{116,118}$Te. 
These mass parameters reflect the topology of their 
PESs, which exhibit two minima and pronounced 
$\gamma$ softness in the case of $^{116}$Te. 

The present study suggests a coexistence of 
an oblate and a near prolate shapes in mid-shell
Te isotopes, as expected experimentally, and provides 
theoretical predictions on these nuclei which are 
experimentally of much interest. 
From the theoretical point of view, 
the deficiency of the employed model in describing the 
correct location of the $0^+_2$ level in the mid-shell 
region stimulates further theoretical investigations 
to reveal possible sources of the uncertainties, 
which are in the present case rooted in the nature 
of the EDF, the SCMF method, the parameters or the 
degrees of freedom involved 
in the QCH.

\acknowledgments
The authors thank Kenichi Yoshida for a valuable discussion.

\bibliography{refs}

\end{document}